\documentclass[twoside,english,american,romanappendices,draftclsnofoot,onecolumn,12pt]{IEEEtran}
\usepackage[T1]{fontenc}
\usepackage[latin9]{inputenc}
\usepackage{float}
\usepackage{mathtools}
\usepackage{amsmath}
\usepackage{amsthm}
\usepackage{amssymb}
\usepackage{graphicx}

\makeatletter

\floatstyle{ruled}
\newfloat{algorithm}{tbp}{loa}
\providecommand{\algorithmname}{Algorithm}
\floatname{algorithm}{\protect\algorithmname}

\theoremstyle{plain}
\newtheorem{thm}{\protect\theoremname}
\theoremstyle{remark}
\newtheorem{rem}[thm]{\protect\remarkname}
\theoremstyle{plain}
\newtheorem{lem}[thm]{\protect\lemmaname}

\usepackage{amsfonts}
\usepackage{amsthm}
\usepackage{mathtools}
\usepackage{cite}
\usepackage{array}
\usepackage{algorithm}
\usepackage{algorithmic}
\usepackage{subfigure}

\DeclareMathOperator{\tr}{tr}

\DeclareMathOperator*{\st}{subject~to}
\DeclareMathOperator*{\maxi}{maximize}
\DeclareMathOperator*{\mini}{minimize}

\newcommand{\herm}{^{\mbox{\scriptsize H}}}
\newcommand{\trans}{^{\mbox{\scriptsize T}}}
\renewcommand{\Im}{\mathrm{Im}}
\renewcommand{\Re}{\mathrm{Re}}

\let\rem\@undefined
\theoremstyle{remark}
\newtheorem{rem}{\protect\remarkname}
\allowdisplaybreaks

\makeatother

\makeatother

\usepackage{babel}
\addto\captionsamerican{\renewcommand{\lemmaname}{Lemma}}
\addto\captionsamerican{\renewcommand{\remarkname}{Remark}}
\addto\captionsamerican{\renewcommand{\theoremname}{Theorem}}
\addto\captionsenglish{\renewcommand{\algorithmname}{Algorithm}}
\addto\captionsenglish{\renewcommand{\lemmaname}{Lemma}}
\addto\captionsenglish{\renewcommand{\remarkname}{Remark}}
\addto\captionsenglish{\renewcommand{\theoremname}{Theorem}}
\providecommand{\lemmaname}{Lemma}
\providecommand{\remarkname}{Remark}
\providecommand{\theoremname}{Theorem}

\begin{document}
\title{Noncoherent Joint Transmission Beamforming for Dense Small Cell Networks:
Global Optimality, Efficient Solution and Distributed Implementation}
\author{Quang-Doanh Vu,~\IEEEmembership{Member, IEEE}, Le-Nam Tran,~\IEEEmembership{Senior Member, IEEE},
and Markku Juntti,~\IEEEmembership{Fellow, IEEE}\thanks{This work was supported in part by the Academy of Finland under the
projects \textquotedblleft Flexible Uplink-Downlink Resource Management
for Energy and Spectral Efficiency Enhancing in Future Wireless Networks
(FURMESFuN)\textquotedblright{} under Grant 310898, and \textquotedblleft 6Genesis
Flagship\textquotedblright{} under Grant 318927. This publication
has emanated from research supported in part by a Grant from Science
Foundation Ireland under Grant number 17/CDA/4786.}\thanks{Quang-Doanh Vu, and Markku Juntti are with Centre for Wireless Communications,
University of Oulu, FI-90014, Finland. Email: \{doanh.vu, markku.juntti\}@oulu.fi.}\thanks{L.-N. Tran is with School of Electrical and Electronic Engineering,
University College Dublin, Ireland. Email: nam.tran@ucd.ie).}\thanks{Part of this work was presented at IEEE International Conference on
Acoustics, Speech and Signal Processing (ICASSP 2019), Brighton, United
Kingdom, May 12-17, 2019 \cite{icassp2019}.}}
\maketitle
\begin{abstract}
We investigate the coordinated multi-point noncoherent joint transmission
(JT) in dense small cell networks. The goal is to design beamforming
vectors for macro cell and small cell base stations (BSs) such that
the weighted sum rate of the system is maximized, subject to a total
transmit power at individual BSs. The optimization problem is inherently
nonconvex and intractable, making it difficult to explore the full
potential performance of the scheme. To this end, we first propose
an algorithm to find a globally optimal solution based on the generic
monotonic branch reduce and bound optimization framework. Then, for
a more computationally efficient method, we adopt the inner approximation
(InAp) technique to efficiently derive a locally optimal solution,
which is numerically shown to achieve near-optimal performance. In
addition, for decentralized networks such as those comprising of multi-access
edge computing servers, we develop an algorithm based on the alternating
direction method of multipliers, which distributively implements the
InAp-based solution. Our main conclusion is that the noncoherent JT
is a promising transmission scheme for dense small cell networks,
since it can exploit the densitification gain, outperforms the coordinated
beamforming, and is amenable to distributed implementation.
\end{abstract}


\section{Introduction}

The rapid growth in the number and the diverse requirements of wireless
communications applications has presented the paramount challenge
of wirelessly transmitting huge volumes of data for the upcoming mobile
networks. It is predicted that the total mobile traffic will be five
times higher by 2023 compared to 2018 \cite{eric_report2018_Nov}.
In addition, the next generation of mobile networks is going to introduce
various service categories to support diverse communication requirements,
e.g., enhanced mobile broadband, massive machine type communications
(mMTC), and ultra-reliable low-latency communications (URLLC) \cite{3gpp5G}.
Dense small cell deployment is one promising technology to enable
the new services \cite{5gdense2015mag}. With densification of low-cost
base stations (BSs), the existing spectrum is exploited efficiently
by the spatial reuse, and the energy efficiency is enhanced due to
the short-range wireless transmission \cite{smallcellNetw}. Furthermore,
the proximity of the cells to the users can support low latency services
as well as guarantee quality of experience (QoE) \cite{smallcellchallenge}.

Designing radio access networks (RANs) has been progressing significantly
during the recent few years. The centralized RAN (CRAN) architecture
moves the baseband processing functionalities of the conventional
base station (BS) to a central location called baseband unit (BBU)
pool \cite{CloudMag2014,cranericson}. This fully centralized architecture
exploits powerful cloud computing capabilities for resource management.
However, it requires local information, e.g., channel state information
(CSI), to be gathered at the centralized BBU pool \cite{cranPoor2015},
which needs a significant cost invested in the fronthaul network and
might result in high latency\cite{MEC_smallcellFor}. In order to
overcome these shortcomings, the concept of multi-access edge computing
(MEC), named by the European Telecommunications Standards Institute
(ETSI), has recently been introduced \cite{MEC_ETSI,MEC_mag2017,MEC_smallcellFor,MEC_survey,MECericson,MEC_ETSI_2015}.
The technology deploying the computing, storage and networking resources
(called MEC servers) across networks allows data to be stored and
processed locally \cite{MEC_smallcellFor,MEC_ETSI_2015,MECericson}.
As such, the networks with MEC technique are sort of decentralized
architecture \cite{MEC_survey,MEC_mag2017}. In dense small cell networks,
MEC servers could be co-located with selected small cell BSs \cite{MEC_smallcellFor}.

In dense small cell networks, the BSs are close to each other. Thus,
efficiently managing inter-cell interference is apparently one of
the keys for a successful implementation\cite{powercontrl}. A common
approach is to use coordinated multi-point (CoMP) strategies. The
simplest form of CoMP is coordinated beamforming (CB) where a specific
user receives data from only one BS, while the interference caused
by other BSs is reduced by the cooperation of the involved BS \cite{COMP:Lee2012}.
The most advanced CoMP strategy is \emph{coherent} joint transmission
(JT) where data for a user is available at multiple BSs, and the BSs
collaborate to create a large virtual multiple-input multiple-output
(MIMO) system in order to maximally exploit array gain \cite{mlticellMIMO_WeiYu2010}.
However, coherent JT requires strict synchronization among BSs (0.5
microsecond accuracy \cite{cransurvey2015}). This requirement remains
a main challenge for a practical implementation of coherent JT \cite{magComp2016},
even in centralized architecture systems (i.e., CRAN) where the synchronization
is improved (compared to the conventional RAN) \cite{cransurvey2015}.
Recently, \emph{noncoherent joint transmission} has received growing
attention \cite{Small_cell_MassMIMO,noncoherentloadbalance,noncoherentJSAC,noncoherentwcom2016},
since it requires not as strict synchronization accuracy as compared
to coherent JT \cite{Small_cell_MassMIMO,noncoherentloadbalance,noncoherentwcom2016,noncoherentJSAC}.
We note that the term `noncoherent' is used herein in the context
of joint transmission referring to the signal processing coordination
among BSs. It does not refer to the classical notions of noncoherent
communications or data detection, in which neither the carrier phase
is available to the receiver \cite{DiProakis}, nor the instantaneous
channel is known at the receiver \cite{TseNoncoherent,MoserNoncoherent}.
In our paper, noncoherent JT refers to the scenario where users still
receive data from multiple BSs, but the data is encoded independently
at individual BSs \cite{Small_cell_MassMIMO}, and users apply successive
interference cancellation to decode its information where the information
from a BS is decoded with the signal from the remaining BSs is treated
as noise \cite{noncoherentJSAC}. As such, noncoherent JT is expected
to require BS synchronization at the same level as CB (3 microsecond
accuracy \cite{cransurvey2015}).

\subsection{Contributions}

The above discussion motivates us to investigate the achievable performance
of the noncoherent JT technique in the context of dense small cell
networks where the BSs collaborate to serve a set of users. The target
is to design beamforming vectors at BSs so that the weighted sum rate
(WSR) is maximized under the constraints on maximum transmit power
at each of the BSs. We consider the WSR as the objective to be maximized,
because it is general enough to encompass other performance measures
such as spectral efficiency and the guaranteed quality of services
for the users (via appropriate weights) as special cases \cite{wsrjsac2006}.
The contributions of this paper are as follows:
\begin{itemize}
\item \emph{Globally optimal solution}: We first find the optimal beamforming
vectors to fully understand the potential performance of the noncoherent
JT. The nonconvexity and intractability of many design problems related
to the WSR maximization have been widely known in the literature \cite{tomLuocomplexity2008}.
To this end, we develop an algorithm based on the branch reduce and
bound (BRnB) monotonic optimization framework which globally solves
the considered problem \cite{tuy2005monotonic}.
\item \emph{Computationally efficient solution}: We develop a low-complexity
suboptimal iterative method based on the well-known inner approximation
(InAp) framework \cite{Inner_Approximation1978,ABeck:SCAmath:2010},
which is provably convergent and efficiently solves the WSR maximization
problem. In each iteration, only a conic quadratic program (CQP) needs
to be solved. Also, we numerically demonstrate the fast convergence
and the near-optimal performance of the suboptimal solution.
\item \emph{Distributed implementation}: We subsequently develop a distributed
implementation of the efficient solution for decentralized architecture
in dense small cell networks deploying the MEC servers. Particularly,
motivated by the appreciated success of the alternating direction
method of multipliers (ADMM) in designing distributed algorithms reported
in recent publications \cite{giangtwcom2017,doanhTSP,ADMM_boyd,distadmm2012},
we rely on this mathematical tool to decompose the convex approximated
problems (i.e., the CQP) obtained by the InAp-based method into subproblems,
which can be solved locally at the MEC servers. As such, the beamforming
vectors can be computed at the MEC servers using local information.
\end{itemize}
We provide extensive numerical results to evaluate the efficiency
of the proposed methods. In particular, we conclude that the noncoherent
JT is suitable for the decentralized architecture dense small cell
networks due to the fact that it has the ability of exploiting densification
gain, and is convenient for being implemented distributively.

\subsection{Related Works}

There is a large portion of related works in the subject of small
cell networks but mostly focusing on coherent JT. For example, the
authors in \cite{EE_COMP_Het_2014} and \cite{doanhTCOM} designed
precoding for minimizing power and maximizing energy efficiency, respectively.
In CRAN based networks, coherent JT was considered in \cite{CranTuyen},
\cite{cranNgo}, \cite{cranLau}, and \cite{cranPhuong} for WSR maximization,
energy efficiency maximization, power minimization, and multi-objective
of spectral and power efficiency maximization, respectively. These
previous works implicitly assume a strict requirement on network synchronization.

The CB has been investigated extensively. The common approaches for
the WSR problem in CB systems include weighted minimum mean square
error (WMMSE) \cite{WMMSE2011} and InAp (or successive convex approximation)
\cite{Nam:WSRMISO:2012}, which were numerically shown to achieve
near-optimal performance \cite{Nam:WSRMISO:2012}. We will see in
the next section that the numerator of the signal-to-interference-plus-noise
ratio (SINR) expression of the noncoherent JT is a sum of multiple
quadratic functions, which is different from the SINR expression of
the CB. Consequently, the solutions developed for CB systems are not
readily suitable for noncoherent JT. Particularly, it remains to be
seen how the WMMSE can be applied to the noncoherent JT, because the
approach of introducing the auxiliary variables in the CB is no longer
useful \cite{noncoherentJSAC}. Also, the InAp-based solution developed
in \cite{Nam:WSRMISO:2012} cannot be directly applied to the noncoherent
JT, since it was derived based on the phase rotation technique, which
does not lead to a tractable formulation in the noncoherent JT context.
The zero forcing (ZF) technique can be applied to the noncoherent
JT \cite{noncoherentwcom2016}, but ZF may be infeasible for dense
small cell networks, because small cell BSs are usually equipped with
a few antennas. Furthermore, a user in a CB system receives desired
signals from only one BS while in the noncoherent JT, it receives
desired signals from multiple BSs. Thus, the existing distributed
algorithms for the CB cannot be straightforwardly applied to the noncoherent
JT.

Noncoherent JT has received growing attention, since it requires less
strict network synchronization accuracy compared to the coherent counterpart
\cite{Small_cell_MassMIMO,noncoherentloadbalance,noncoherentwcom2016,noncoherentJSAC}.
In \cite{Small_cell_MassMIMO}, beamforming vectors at the BSs were
designed for minimizing the power consumption subject to users' minimum
data rate. In \cite{noncoherentwcom2016}, noncoherent JT was studied
for the two problems: (i) power minimization subject to users' minimum
data rate, and (ii) weighted max-min fairness. Therein, each BS is
equipped with a massive number of antennas and simple beamforming
schemes, i.e., ZF and maximum ratio transmission (MRT), are used.
As discussed above, ZF is not a feasible approach for small cell systems.
It is noted that, different from the power minimization problem, applying
MRT for a WSR maximizing scheme does not lead to a tractable problem.
Noncoherent JT design for minimizing weighted power consumption with
imperfect channel state information was considered in \cite{noncoherentloadbalance}.
A heuristic beamforming design for maximizing the WSR under the limited
fronthaul capacity for CRAN networks was proposed in \cite{noncoherentJSAC},
which showed that noncoherent JT might outperform coherent JT in the
regime of low fronthaul capacity. Generally, for power minimization
problems, the optimal beamforming vectors for noncoherent JT can be
exactly found since their semidefinite relaxation (SDR) versions are
tight and convex \cite{Small_cell_MassMIMO}. However, for the WSR
maximization, its SDR is still intractable.

In summary, the full potential performance and useful insights into
the design of noncoherent JT in dense small cell networks in terms
of WSR maximization has not been previously well studied. Moreover,
there is a lack of an efficient distributed algorithm implementing
noncoherent JT in decentralized architecture networks. These objectives
are the main focus in this paper.

\subsection{Organization and Notations}

The rest of the paper is organized as follows. Section \ref{sec:System-Model}
describes the system model and the problem formulation of designing
noncoherent JT beamforming for maximizing WSR. Section \ref{sec:Optimal-Solution-to}
presents a globally optimal solution of the problem. An efficient
solution is provided in Section \ref{sec:Efficient-Solutions} followed
by its distributed implementation presented in Section \ref{sec:Distributed-Algorithm-via}.
Numerical results and discussions are provided in Section \ref{sec:Numerical-Results}.
Finally, Section \ref{sec:Conclusion} concludes the work.

\emph{Notation}s: Bold lower and upper case letters represent vectors
and matrices, respectively; $||\cdot||_{p}$ represents the $\ell_{p}$
norm; $|\cdot|$ is the absolute value of the argument; $\mathbb{C}^{x\times y}$
represents the space of complex matrices of dimensions given in the
superscript; $\mathbb{S}_{+}^{x}$ denotes the space of symmetric
positive semidefinite matrices; $\mathcal{CN}(0,a)$ denotes a complex
Gaussian random variable with zero mean and variance $a$; $\Re\{\cdot\}$
represents real part of the argument. Notation $\mathbf{e}_{i}$ denotes
the $i\textrm{th}$ conventional basis vector, i.e., the vector such
that $e_{i}=1$ and $e_{j}=0,\forall j\neq i$. $\mathbf{X}\trans$
and $\mathbf{X}\herm$ stand for the transpose and the Hermitian transpose
of $\mathbf{X}$, respectively. We use \textquotedblleft MATLAB notation\textquotedblright{}
$\textrm{blkdiag}\{\cdot\}$ which represents block diagonal matrix.

\section{System Model\label{sec:System-Model}}

\subsection{Signal Transmission}

\begin{figure}
\begin{centering}
\includegraphics[width=0.45\columnwidth]{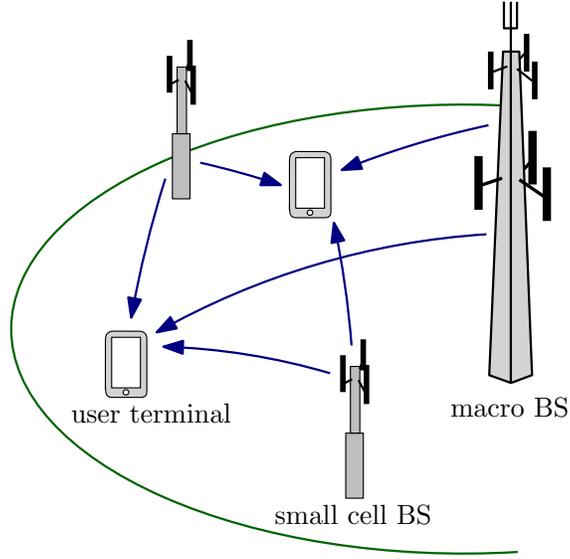}
\par\end{centering}
\caption{Small cell deployment system model.}
 \label{fig:systemmodel}
\end{figure}

We consider a region covered by a macro cell BS and a set of $K$
small cell BSs shown in Fig.\ \ref{fig:systemmodel}. Let us denote
by $\mathcal{K}=\{1,2,...,K+1\}$ the set of all BSs where $\{1\}$
refers to the macro BS and the rest the small cell BSs. BS $k$ is
equipped with $M_{k}$ antennas. The BSs simultaneously serve a set
of $N$ single-antenna users, denoted by $\mathcal{N}=\{1,2,...,N\}$
under the same frequency band. Let $\mathrm{B}_{k}$ and $\mathrm{U}_{i}$
denote BS $k$ and user $i$, respectively. Herein, we assume that
the BSs collaborate using nonconherent JT, i.e., the information for
a specific user is encoded independently at individual BSs \cite{Small_cell_MassMIMO}.
Particularly, let $s_{ik}$ and $\mathbf{v}_{ik}\in\mathbb{C}^{M_{k}\times1}$
be the normalized symbol and the beamforming vector at $\mathrm{B}_{k}$
for $\mathrm{U}_{i}$, respectively. Let $\mathbf{h}_{ik}\in\mathbb{C}^{1\times M_{k}}$
(row vector) be the channel between $\mathrm{B}_{k}$ and $\mathrm{U}_{i}$,
which is assumed to be perfectly known. The signal received at $\mathrm{U}_{i}$
under the assumption of flat channels is given by 
\begin{eqnarray}
r_{i} & = & {\textstyle \sum_{k\in\mathcal{K}}}\mathbf{h}_{ik}\mathbf{v}_{ik}s_{ik}+{\textstyle \sum_{k\in\mathcal{K}}\sum_{\substack{j\in\mathcal{N}\setminus\{i\}}
}}\mathbf{h}_{ik}\mathbf{v}_{jk}s_{jk}+z_{i}\label{eq:SigReceive}
\end{eqnarray}
where $z_{i}\thicksim\mathcal{CN}(0,\sigma_{i}^{2})$ is the additive
white Gaussian noise. The first and second sum in the right side of
\eqref{eq:SigReceive} are the desired signal and the interference,
respectively. The users are assumed to use successive interference
cancellation technique to detect its own signal and treat signal of
other users as noise. Thus the \emph{effective} (or aggregated) SINR
at $\mathrm{U}_{i}$ can be written as \cite{Small_cell_MassMIMO}\footnote{We note that the decoding order has no impact on $\gamma_{i}$ \cite{noncoherentJSAC}.}
\begin{equation}
\gamma_{i}(\{\mathbf{v}_{ik}\})=\frac{\sum_{k\in\mathcal{K}}|\mathbf{h}_{ik}\mathbf{v}_{ik}|^{2}}{\sum_{k\in\mathcal{K}}\sum_{\substack{j\in\mathcal{N}\setminus\{i\}}
}|\mathbf{h}_{ik}\mathbf{v}_{jk}|^{2}+\sigma_{i}^{2}}.\label{eq:sinrsmall}
\end{equation}
We note that $\{\gamma_{i}(\{\mathbf{v}_{ik}\})\}_{i\in\mathcal{N}}$
are achieved without phase synchronization between BSs. We also remark
that $\{\gamma_{i}\}_{i\in\mathcal{N}}$ are the aggregated instantaneous
SINR, i.e., the total information received at $\mathrm{U}_{i}$ is
$\log(1+\gamma_{i})$ \cite{Small_cell_MassMIMO}. The reader is referred
to \cite{noncoherentJSAC} for the derivation of $\gamma_{i}(\{\mathbf{v}_{ik}\})$.

\subsection{Problem Formulation}

We aim at designing beamforming vectors $\{\mathbf{v}_{ik}\}_{i,k}$
to maximize the WSR under the constraints of the transmit power budget
at the BSs. Mathematically, the problem reads \begin{subequations}\label{eq:wsr_prob}
\begin{align}
\underset{\{\mathbf{v}_{ik}\}}{\maxi} & \;{\textstyle \sum_{i\in\mathcal{N}}}w_{i}\log(1+\gamma_{i}(\{\mathbf{v}_{ik}\}))\label{eq:GeneralObj}\\
\st & \;{\textstyle \sum_{i\in\mathcal{N}}}\mathbf{v}_{ik}\herm\mathbf{v}_{ik}\leq P_{k},\forall k\in\mathcal{K}\label{eq:powerconst}
\end{align}
\end{subequations} where $w_{i}>0$ is the priority of $\mathrm{U}_{i}$,
and $P_{k}$ is the maximum transmit power available to $\mathrm{B}_{k}$.
Note that the SINR in \eqref{eq:sinrsmall} is nonconvex with $\{\mathbf{v}_{ik}\}_{i,k}$,
which makes problem \eqref{eq:wsr_prob} intractable \cite{Small_cell_MassMIMO}.

\subsection{Centralized and Distributed Architectures of Small Cell Networks}

\begin{figure}
\begin{centering}
\includegraphics[width=1\columnwidth]{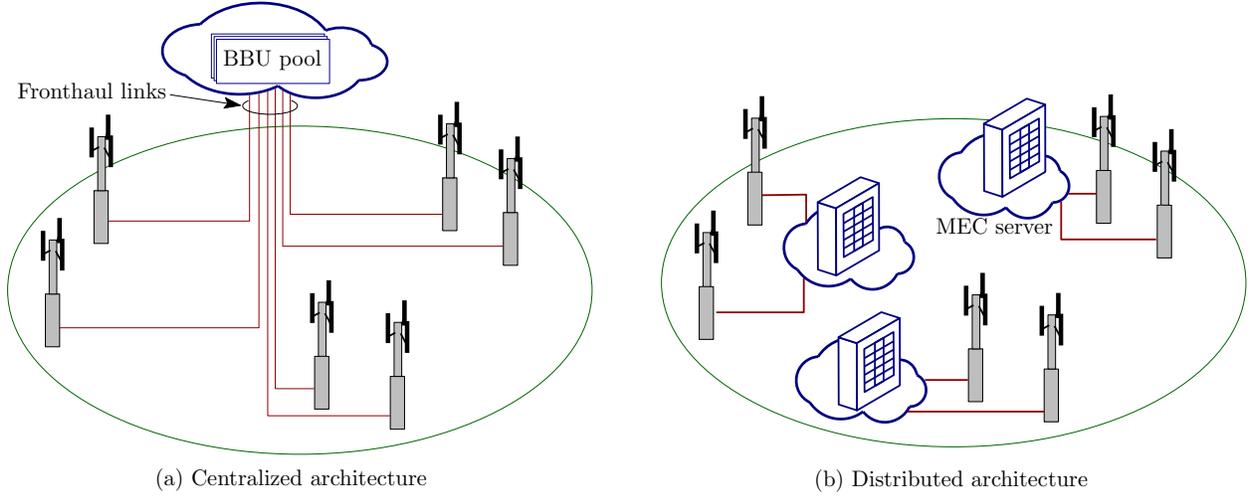}
\par\end{centering}
\caption{Centralized and distributed network architectures.}
 \label{fig:networkArchi}
\end{figure}

The noncoherent JT technique can be deployed on the centralized or
distributed platforms as illustrated in Fig.\ \ref{fig:networkArchi}.
The centralized one mainly refers to the CRANs where the baseband
processing functionalities of the BSs are centralized at the BBU pool
\cite{CloudMag2014,cranericson}. It requires channel vector $\mathbf{h}_{ik}$
to be gathered at the BBU pool for all $i\in\mathcal{N},k\in\mathcal{K}$,
where all beamforming vectors $\{\mathbf{v}_{ik}\}_{i\in\mathcal{N},k\in\mathcal{K}}$
are calculated. After baseband processing, the signal is sent to the
BSs (remote radio head to be precise) via fronthaul links before being
transmitted to users via the wireless interface. We herein focus on
the performance of noncoherent JT over wireless channels. Thus, we
suppose that the capacity of the fronthaul links are sufficiently
large and not forming the bottleneck. 

Distributed platform refers to the MEC architecture \cite{MEC_ETSI,MEC_mag2017,MEC_smallcellFor,MEC_survey,MECericson,MEC_ETSI_2015},
which is introduced to support low-latency services. Here, multiple
MEC servers can be deployed across the networks to store and process
data locally. In dense small cell networks, baseband processing function
of one or several BSs located close to each other can be gathered
at a MEC server to exploit the server's computing capacity. A MEC
server can be physically co-located with a selected BS \cite{MEC_smallcellFor}.
To efficiently manage interference of dense small cells, the MEC servers
should cooperate with each other in baseband processing. 

We herein consider both centralized and distributed network architectures.
For the latter, we suppose that there is a set of $D$ MEC servers
denoted by $\mathcal{D}=\{1,...,D\}$. MEC server $d$ handles the
baseband processing functionality of a set of BSs denoted by $\mathcal{K}_{d}$,
$\mathcal{K}_{d}\subset\mathcal{K}$. We also suppose that each of
the MEC servers can serve multiple BSs, and each of the BSs is served
by only one MEC server, i.e., $|\mathcal{K}_{d}|\geq1$, $\mathcal{K}_{d}\cap\mathcal{K}_{d'}=\emptyset$,
$\forall d\neq d'$, and $\underset{d\in\mathcal{D}}{\cup}\mathcal{K}_{d}=\mathcal{K}$.
Thus, beamforming vectors $\{\mathbf{v}_{ik}\}_{i\in\mathcal{N},k\in\mathcal{K}_{d}}$
can be locally computed at MEC server $d$ if the local CSI, i.e.,
$\{\mathbf{h}_{ik}\}_{i\in\mathcal{N},k\in\mathcal{K}_{d}}$, is available.

\section{Globally Optimal Solution to \eqref{eq:wsr_prob}: A BRnB Algorithm\label{sec:Optimal-Solution-to}}

In this section, we develop an algorithm which globally solves \eqref{eq:wsr_prob}
based on the BRnB monotonic optimization framework \cite{tuy2005monotonic}.
The purpose of finding a global solution to \eqref{eq:wsr_prob} is
twofold: (i) exploring the best achievable performance of the noncoherent
JT technique in small cell networks, and (ii) benchmarking against
the efficient suboptimal solution to be presented. We start by rewriting
\eqref{eq:wsr_prob} as\begin{subequations}\label{eq:wsr_prob-1}
\begin{align}
\underset{\{\mathbf{v}_{ik}\},\mathbf{r}\in\mathbb{R}_{+}^{N\times1}}{\maxi} & \;f(\mathbf{r})\triangleq{\textstyle \sum_{i\in\mathcal{N}}}w_{i}r_{i}\label{eq:GeneralObj-2}\\
\st & \;\log(1+\gamma_{i}(\{\mathbf{v}_{ik}\}))\geq r_{i},\forall i\in\mathcal{N}\label{eq:lograte}\\
 & \;{\textstyle \sum_{i\in\mathcal{N}}}\mathbf{v}_{ik}\herm\mathbf{v}_{ik}\leq P_{k},\forall k\in\mathcal{K}\label{eq:powerconst-2}
\end{align}
\end{subequations}Problem \eqref{eq:wsr_prob-1} is the epigraph
representation of \eqref{eq:wsr_prob}, thus the two problems are
equal in the sense of optimal solutions.

The following properties make \eqref{eq:wsr_prob-1} suitable for
applying BRnB. First, since $w_{i}\geq0$, objective function $f(\mathbf{r})$
is monotonically increasing with respect to $\mathbf{r}$, i.e., $f(\mathbf{r}')\leq f(\mathbf{r})$
for all $\mathbf{r}'\leq\mathbf{r}$ where the inequality is understood
element wise. Second, let $\check{r}_{i}=0$ and $\hat{r}_{i}=\log\Bigl(1+\frac{\sum_{k\in\mathcal{K}}P_{k}\mathbf{h}_{ik}\mathbf{h}_{ik}\herm}{\sigma_{i}^{2}}\Bigr),\,\forall i\in\mathcal{N}$,
then the box $[\check{\mathbf{r}},\hat{\mathbf{r}}]$ contains all
feasible points $\mathbf{r}$ in \eqref{eq:wsr_prob-1}. Third, the
set $\mathcal{S}\triangleq\{\mathbf{r}\in\mathbb{R}_{+}^{N\times1},\eqref{eq:lograte},\eqref{eq:powerconst-2}\}$
is normal compact in $[\check{\mathbf{r}},\hat{\mathbf{r}}]$, i.e.,
if $\mathbf{r}'\in\mathcal{S}$, then $[\check{\mathbf{r}},\mathbf{r}']\subset\mathcal{S}$.

A BRnB monotonic algorithm is an iterative procedure comprising of
three main steps called branching, reduction and bounding. In what
follows, we present the details of these steps customized to \eqref{eq:wsr_prob-1}.

Let us consider iteration $t$ and define some notations. Let $\mathtt{lb}_{\mathrm{best}}$
and $\mathbf{r}_{\mathrm{best}}$ denote the current best lower bound
and the feasible point achieving $\mathtt{lb}_{\mathrm{best}}$, respectively.
Let $\mathcal{Q}$ be the set of candidate boxes. Let $\mathtt{ub}(R)$
denote an upper bound on $f(\mathbf{r})$ for feasible $\mathbf{r}$
in box $R$ (i.e. $\mathbf{r}\in\mathcal{S}\cap R$). The way to calculate
$\mathtt{ub}(R)$ is presented in the bounding step.

\subsection*{Branching}

In this step, a box in $\mathcal{Q}$ is picked and then divided into
two smaller boxes. Specifically, let $R^{(t)}=[\underline{\mathbf{r}},\bar{\mathbf{r}}]$
denote the chosen box which has the largest upper bound compared to
those in $\mathcal{Q}^{(t)}$, i.e.,
\begin{equation}
R^{(t)}=\underset{R\in\mathcal{Q}}{\textrm{argmax}}\;\mathtt{ub}(R).\label{eq:select_box}
\end{equation}
 Box $R^{(t)}$ is divided into two boxes $R_{1}^{(t)}=[\underline{\mathbf{r}}_{1},\bar{\mathbf{r}}_{1}]$
and $R_{2}^{(t)}=[\underline{\mathbf{r}}_{2},\bar{\mathbf{r}}_{2}]$,
$R_{1}^{(t)}\cup R_{2}^{(t)}=R^{(t)}$, which are determined as 
\begin{align}
\underline{\mathbf{r}}_{1}=\underline{\mathbf{r}}+0.5(\bar{r}_{i}-\underline{r}_{i})\mathbf{e}_{i},\;\bar{\mathbf{r}}_{1}=\bar{\mathbf{r}},\;\underline{\mathbf{r}}_{2}=\underline{\mathbf{r}},\;\bar{\mathbf{r}}_{2}=\bar{\mathbf{r}}-0.5(\bar{r}_{i}-\underline{r}_{i})\mathbf{e}_{i}\label{eq:branchbox}
\end{align}
where $i\in\mathcal{N}$. The upper bound of the new boxes are $\mathtt{ub}(R_{1}^{(t)})=\mathtt{ub}(R)$
, and $\mathtt{ub}(R_{2}^{(t)})=\textrm{min}\{\mathtt{ub}(R),f(\bar{\mathbf{r}}_{2})\}$
.
\begin{rem}
\emph{\label{rem:Branchingrule}{[}Branching with weighted edges{]}}
Commonly, the longest edge of $R^{(t)}$ is selected to be branched,
i.e., $i=\underset{i'\in\mathcal{N}}{\textrm{argmax}}\;(\bar{r}_{i'}-\underline{r}_{i'})$
\cite{EE_Osk,robustBjon}. For the WSR problem \eqref{eq:wsr_prob-1},
we propose the rule of choosing the longest weighted edge for branching,
i.e., $i=\underset{i'\in\mathcal{N}}{\textrm{argmax}}\;w_{i'}(\bar{r}_{i'}-\underline{r}_{i'})$.
The proposed rule is numerically shown to accelerate the convergence
(cf. Fig.\ \ref{fig:convergeBnB}).
\end{rem}

\subsection*{Reduction}

The reduction step is to remove the parts of a box which is guaranteed
not to contain an optimal solution. Let us consider the box $R=[\underline{\mathbf{r}},\bar{\mathbf{r}}]$
which has $\mathtt{ub}(R)\geq\mathtt{lb}_{\mathrm{best}}$. The reduction
of $R$ is denoted by $\mathtt{red}(R)=[\underline{\mathbf{r}}',\bar{\mathbf{r}}']$,
where $\underline{\mathbf{r}}\leq\underline{\mathbf{r}}'$ and $\bar{\mathbf{r}}\geq\bar{\mathbf{r}}'$
express as 
\begin{equation}
\underline{\mathbf{r}}'=\bar{\mathbf{r}}-{\textstyle \sum_{i=1}^{N}}\beta_{i}(\bar{r}_{i}-\underline{r}_{i})\mathbf{e}_{i},\;\bar{\mathbf{r}}'=\underline{\mathbf{r}}'+{\textstyle \sum_{i=1}^{N}}\alpha_{i}(\bar{r}_{i}-\underline{r}'_{i})\mathbf{e}_{i}
\end{equation}
where 
\begin{equation}
\begin{array}{c}
\beta_{i}=\textrm{argmax}\{\beta|\beta\in[0,1],f(\bar{\mathbf{r}}-\beta(\bar{r}_{i}-\underline{r}_{i})\mathbf{e}_{i})\geq\mathtt{lb}_{\mathrm{best}}\}\\
\alpha_{i}=\textrm{argmax}\{\alpha|\alpha\in[0,1],f(\underline{\mathbf{r}}'+\alpha(\bar{r}_{i}-\underline{r}'_{i})\mathbf{e}_{i})\leq\mathtt{ub}(R)\}.
\end{array}\label{eq:cuts}
\end{equation}
It is guaranteed that any optimal solution contained in $R$ is in
$\mathtt{red}(R)$. Indeed, with the $\{\alpha_{i}\}$ and $\{\beta_{i}\}$
in \eqref{eq:cuts}, we have $f(\underline{\mathbf{r}}')\leq\mathtt{lb}_{\mathrm{best}}$
and $f(\bar{\mathbf{r}}')\geq\mathtt{ub}(R)$. This means only the
points $\mathbf{r}\in R$ with $f(\mathbf{r})<\mathtt{lb}_{\mathrm{best}}$
or $f(\mathbf{r})>\mathtt{ub}(R)$ are removed. In fact, \eqref{eq:cuts}
can be rewritten in the following closed-form expressions 
\[
\beta_{i}=\textrm{min}\Bigl\{1,\frac{f(\bar{\mathbf{r}})-\mathtt{lb}_{\mathrm{best}}}{w_{i}(\bar{r}_{i}-\underline{r}_{i})}\Bigr\},\,\alpha_{i}=\textrm{min}\Bigl\{1,\frac{\mathtt{ub}(R)-f(\underline{\mathbf{r}}')}{w_{i}(\bar{r}_{i}-\underline{r}'_{i})}\Bigr\}.
\]

\subsection*{Bounding}

This operation is to improve $\mathtt{lb}_{\mathrm{best}}$ and the
upper bound of a box, from which the boxes containing no feasible
point whose objective value is larger than the current $\mathtt{lb}_{\mathrm{best}}$
are removed. Let us consider the box $R=[\underline{\mathbf{r}},\bar{\mathbf{r}}]$
where $\underline{\mathbf{r}}$ is feasible (otherwise, $R$ does
not contain any feasible point, and thus should be removed). Let $\boldsymbol{\phi}=\frac{\bar{\mathbf{r}}-\underline{\mathbf{r}}}{||\bar{\mathbf{r}}-\underline{\mathbf{r}}||_{2}}$,
$\delta_{\mathrm{low}}=\textrm{argmax}\{\delta|(\underline{\mathbf{r}}+\delta\boldsymbol{\phi})\in\mathcal{S}\}$,
and $\delta_{\mathrm{up}}=\textrm{argmin}\{\delta|(\underline{\mathbf{r}}+\delta\boldsymbol{\phi})\in[\check{\mathbf{r}},\hat{\mathbf{r}}]\setminus\mathcal{S}\}$.
Clearly, the best objective value achieved by the feasible points
in $R$ lies in the segment $[f(\underline{\mathbf{r}}+\delta_{\mathrm{low}}\boldsymbol{\phi}),f(\underline{\mathbf{r}}+\delta_{\mathrm{up}}\boldsymbol{\phi})]$.
Thus, we can update $\mathtt{lb}_{\mathrm{best}}$ and $\mathtt{ub}(R)$
as: $\mathtt{lb}_{\mathrm{best}}=\textrm{max}\{f(\underline{\mathbf{r}}+\delta_{\mathrm{low}}\boldsymbol{\phi}),\mathtt{lb}_{\mathrm{best}}\}$
and $\mathtt{ub}(R)=\textrm{min}\{\textrm{max}\{f(\bar{\mathbf{r}}-(\bar{r}_{i}-\underline{r}_{i}-\delta_{\mathrm{up}}\phi_{i})\mathbf{e}_{i}|i\in\mathcal{N}\},\mathtt{ub}(R)\}$.
The values of $\delta_{\mathrm{low}}$ and $\delta_{\mathrm{up}}$
can be determined by the bisection algorithm over the interval $[0,||\bar{\mathbf{r}}-\underline{\mathbf{r}}||_{2}]$.

\subsubsection*{Checking Feasibility}

From the above discussions, it becomes apparent that checking whether
a given point is feasible or not plays the key role in bounding. Given
$\mathbf{r}$, the feasibility problem is given by\begin{subequations}\label{eq:feasible_prob}
\begin{align}
\tilde{\mathcal{S}}(\mathbf{r})=\mathrm{find} & \;\{\mathbf{v}_{ik}\}\label{eq:GeneralObj-2-1}\\
\st & \;\frac{\sum_{k\in\mathcal{K}}|\mathbf{h}_{ik}\mathbf{v}_{ik}|^{2}}{\sum_{k\in\mathcal{K}}\sum_{\substack{j\in\mathcal{N}\setminus\{i\}}
}|\mathbf{h}_{ik}\mathbf{v}_{jk}|^{2}+\sigma_{i}^{2}}\geq\tilde{r}_{i},\forall i\in\mathcal{N}\label{eq:rate_feas}\\
 & \;{\textstyle \sum_{i\in\mathcal{N}}}\mathbf{v}_{ik}\herm\mathbf{v}_{ik}\leq P_{k},\forall k\in\mathcal{K}\label{eq:power_feas}
\end{align}
\end{subequations}where $\tilde{r}_{i}=\exp(r_{i})-1$. The feasible
set in \eqref{eq:feasible_prob} is nonconvex due to \eqref{eq:rate_feas}.
Different from the coherent JT, using the trick of a phase rotation
here does not lead to a tractable formulation due to the sum of quadratic
functions at numerator in \eqref{eq:rate_feas} \cite{EE_Osk}. However,
problem \eqref{eq:feasible_prob} can be solved exactly using semidefinite
relaxation (SDR). In particular, let us write the SDR of \eqref{eq:feasible_prob}
as \begin{subequations}\label{eq:feasible_prob_sd}
\begin{align}
\tilde{\mathcal{S}}_{\mathrm{SD}}(\mathbf{r})=\mathrm{find} & \;\{\mathbf{V}_{ik}\}\label{eq:GeneralObj-2-1-1-1}\\
\st & \;{\textstyle \sum_{k\in\mathcal{K}}}\mathbf{h}_{ik}\mathbf{V}_{ik}\mathbf{h}_{ik}\herm\geq\tilde{r}_{i}{\textstyle \sum_{k\in\mathcal{K}}\sum_{\substack{j\in\mathcal{N}\setminus\{i\}}
}}\mathbf{h}_{ik}\mathbf{V}_{jk}\mathbf{h}_{ik}\herm+\tilde{r}_{i}\sigma_{i}^{2},\forall i\in\mathcal{N}\label{eq:relaxation1}\\
 & \;{\textstyle \sum_{i\in\mathcal{N}}}\tr(\mathbf{V}_{ik})\leq P_{k},\forall k\in\mathcal{K},\;\mathbf{V}_{ik}\in\mathbb{S}_{+}^{M_{k}},\forall i\in\mathcal{N},\forall k\in\mathcal{K}.\label{eq:relaxation2}
\end{align}
\end{subequations} The relationship between \eqref{eq:feasible_prob}
and \eqref{eq:feasible_prob_sd} in terms of feasibility is stated
in the following lemma.
\begin{lem}
\label{lem:feasibility}For a given point $\mathbf{r}$, the set $\tilde{\mathcal{S}}(\mathbf{r})$
is nonempty if and only if $\tilde{\mathcal{S}}_{\mathrm{SD}}(\mathbf{r})$
is nonempty.
\end{lem}
\begin{IEEEproof}
The proof is given in Appendix \ref{sec:Proof-of-Lemma-feasibility}.
\end{IEEEproof}
The result in Lemma \ref{lem:feasibility} means that the feasibility
of some point $\mathbf{r}$ can be justified via \eqref{eq:feasible_prob_sd}.

\subsection*{The Globally Optimal Algorithm}

The proposed BRnB algorithm for solving \eqref{eq:wsr_prob} is outlined
in Algorithm \ref{alg:glabalsol}. Line \ref{alg:branch} is branching,
line \ref{alg:reduce} is reduction, and lines \ref{alg:bound_be}-\ref{alg:bound_end}
are bounding, as explained above. At the initial stage (line \ref{alg:initial}),
we can randomly generate $\{\mathbf{v}_{ik}\}$ such that \eqref{eq:powerconst}
is satisfied, and then determine $\mathbf{r}_{\mathrm{best}}$ by
letting \eqref{eq:lograte} hold with equality. Notation $\mathtt{UB}$
denotes the current largest upper bound (of the boxes in $\mathcal{Q}$).
Removing boxes that do not contain any optimal solution is shown in
line \ref{alg:removebox}. The stopping criterion in line \ref{alg:stop}
ensures that the output $\mathtt{lb}_{\mathrm{best}}$ is not lower
than $100(1-\epsilon)\%$ of the optimality. To determine beamforming
vectors $\{\mathbf{v}_{ik}^{\star}\}$ achieving $\mathbf{r}_{\mathrm{best}}$,
i.e., line \ref{alg:achieveBeam}, we first solve problem
\begin{equation}
\underset{\{\mathbf{V}_{ik}\}}{\mini}\;{\textstyle \sum_{i\in\mathcal{N}}\sum_{k\in\mathcal{K}}}\tr(\mathbf{V}_{ik})\;\st\textrm{ \{\eqref{eq:relaxation1}, \eqref{eq:relaxation2}\}}
\end{equation}
and denote the obtained solution by $\{\mathbf{V}'_{ik}\}$. If $\mathrm{rank}(\mathbf{V}'_{ik})\leq1$,
we extract $\mathbf{v}_{ik}^{\star}$ via the eigenvalue decomposition
of $\mathbf{V}'_{ik}$. Otherwise, $\mathbf{v}_{ik}^{\star}$ can
be found as the solution to the following CQP derived from \eqref{eq:feasible_prob_sd-1}
(see Appendix \ref{sec:Proof-of-Lemma-feasibility})
\begin{equation}
\underset{||\mathbf{v}||_{2}^{2}\leq\tr(\mathbf{V}'_{ik})}{\maxi}\;\Re\{\mathbf{h}_{ik}\mathbf{v}\}\;\st\;\mathbf{v}\herm\mathbf{h}_{jk}\herm\mathbf{h}_{jk}\mathbf{v}\leq\mathbf{h}_{jk}\mathbf{V}'_{ik}\mathbf{h}_{jk}\herm,\forall j\neq i.
\end{equation}
We now discuss the optimality of Algorithm \ref{alg:glabalsol}. In
particular, let us denote by $f_{\mathrm{opt}}$ the optimal objective
value. Then we have the following lemma.
\begin{lem}
\label{thm:global}Given any $\varepsilon>0$, Algorithm \ref{alg:glabalsol}
guarantees to achieve $\mathtt{UB}-\mathtt{lb}_{\mathrm{best}}<\varepsilon$
where $f_{\mathrm{opt}}\in[\mathtt{lb}_{\mathrm{best}},\mathtt{UB}]$
in a finite number of iterations.
\end{lem}
\begin{IEEEproof}
A proof is provided in Appendix \ref{sec:Proof-of-Theorem}.
\end{IEEEproof}
Since $\mathbf{r}_{\mathrm{best}}$ is feasible, the lemma means that
we can find an $\varepsilon$-approximate optimal solution, i.e. $f_{\mathrm{opt}}-f(\mathbf{r}_{\mathrm{best}})<\varepsilon$,
for any $\varepsilon>0$ after a finite number of iterations. 

The computational complexity of each iteration in Algorithm \ref{alg:glabalsol}
is mainly incurred by solving feasibility problems \eqref{eq:feasible_prob_sd}
at the bounding process. More explicitly, problem \eqref{eq:feasible_prob_sd}
contains $2N\sum_{k}M_{k}^{2}$ real variables, $(N+K)$ constraints
in size 1, and $N$ constraints in size $\sum_{k}2M_{k}$ for each
$k$. So, the \emph{worst-case} of computational cost for solving
\eqref{eq:feasible_prob_sd} is $\mathcal{O}(\sqrt{K+N(1+2\sum_{k}M_{k})}\linebreak4N^{2}(\sum_{k}M_{k}^{2})^{2}(K+N(1+4\sum_{k}M_{k}^{2})))$
\cite{SOCApp_boyd1999}. The number of problems needed to be solved
depends on the bisection accuracy, denoted by $\epsilon_{\textrm{bi}}$,
of determining $\delta_{\mathrm{low}}$ and $\delta_{\mathrm{up}}$,
which is $\mathcal{O}(\log_{2}(1/\epsilon_{\textrm{bi}}))$.

\begin{algorithm}[!t]
\caption{A Branch Reduce and Bound Algorithm to Globally Solve (\ref{eq:wsr_prob})}
\label{alg:glabalsol} \begin{algorithmic}[1]

\STATE \textbf{Initialization}: set $t\coloneqq1$, $\mathcal{Q}\coloneqq[\check{\mathbf{r}},\hat{\mathbf{r}}]$,
$\mathtt{UB}\coloneqq f(\hat{\mathbf{r}})$. Set initial $\mathbf{r}_{\mathrm{best}}$
and $\mathtt{lb}_{\mathrm{best}}=f(\mathbf{r}_{\mathrm{best}})$.
Given accuracy parameter $\epsilon$. \label{alg:initial}

\REPEAT 

\STATE Select $R^{(t)}$ as \eqref{eq:select_box}, then branch $R^{(t)}$
into $R_{1}^{(t)}$ and $R_{2}^{(t)}$ as \eqref{eq:branchbox}. $\mathcal{Q}\coloneqq\mathcal{Q}\setminus R^{(t)}$.\label{alg:branch}

\STATE \textbf{if }$\mathtt{ub}(R_{m}^{(t)})\geq\mathtt{lb}_{\mathrm{best}}$
where $m\in\{1,2\}$, \textbf{then} determine $\mathtt{red}(R_{m}^{(t)})$
and update $\mathcal{Q}\coloneqq\mathcal{Q}\cup\mathtt{red}(R_{m}^{(t)})$
\textbf{end}.\label{alg:reduce}

\STATE \textbf{if }$\mathtt{red}(R_{1}^{(t)})=[\underline{\mathbf{r}}_{1}',\bar{\mathbf{r}}_{1}']$
contains feasible points (check feasibility of $\underline{\mathbf{r}}_{1}'$)
\textbf{then }\label{alg:bound_be}

\STATE $\;$determine $\boldsymbol{\phi}=\frac{\bar{\mathbf{r}}_{1}'-\underline{\mathbf{r}}_{1}'}{||\bar{\mathbf{r}}_{1}'-\underline{\mathbf{r}}_{1}'||_{2}}$,
$\delta_{\mathrm{low}}$ and $\delta_{\mathrm{up}}$, and update $\mathtt{lb}_{\mathrm{best}}$
, $\mathbf{r}_{\mathrm{best}}$, and $\mathtt{ub}(\mathtt{red}(R_{1}^{(t)}))$

\STATE \textbf{else }set $\mathtt{ub}(\mathtt{red}(R_{1}^{(t)}))=0$
\textbf{end}. \label{alg:bound_end}

\STATE Update $\mathcal{Q}\coloneqq\mathcal{Q}\setminus\{R|\mathtt{ub}(R)<\mathtt{lb}_{\mathrm{best}}\}$.\label{alg:removebox}

\STATE Update $\mathtt{UB}\coloneqq\textrm{max}\{\mathtt{ub}(R)|R\in\mathcal{Q}\}$,
$t:=t+1$.

\UNTIL {$\frac{\mathtt{UB}-\mathtt{lb}_{\mathrm{best}}}{\mathtt{lb}_{\mathrm{best}}}\leq\epsilon$}
\label{alg:stop}

\STATE Determine beamforming vectors $\{\mathbf{v}_{ik}^{\star}\}$
which achieve $\mathbf{r}_{\mathrm{best}}$.\label{alg:achieveBeam}

\STATE \textbf{Output}: $\mathbf{r}_{\mathrm{best}}$, $\mathtt{lb}_{\mathrm{best}}$,
and $\{\mathbf{v}_{ik}^{\star}\}$.\end{algorithmic} 
\end{algorithm}

\section{Efficient Solution to \eqref{eq:wsr_prob}: An InAp-based Algorithm\label{sec:Efficient-Solutions}}

The globally optimal algorithm presented in the previous section comes
at a high computational cost, and might be unsuitable for a real-time
practical implementation. In this section, we present a fast converging
and low-complexity solution to \eqref{eq:wsr_prob} based on the InAp
framework, which was inspired by our earlier work in \cite{Nam:WSRMISO:2012}.
To do so, we first transform \eqref{eq:wsr_prob} into an equivalent
form where convexity is easily justified as follows\begin{subequations}\label{eq:EE_prob-1}
\begin{align}
\underset{\{\mathbf{v}_{ik}\},\{\mu_{i}\}}{\maxi} & \;{\textstyle \sum_{i\in\mathcal{N}}}w_{i}\log(1+\mu_{i})\label{eq:GeneralObj-1}\\
\st & \;\frac{{\textstyle \sum_{k\in\mathcal{K}}}|\mathbf{h}_{ik}\mathbf{v}_{ik}|^{2}}{{\textstyle \sum_{k\in\mathcal{K}}\sum_{\substack{j\in\mathcal{N}\setminus\{i\}}
}}|\mathbf{h}_{ik}\mathbf{v}_{jk}|^{2}+\sigma_{i}^{2}}\geq\mu_{i},\forall i\in\mathcal{N},\label{eq:snrconstr}\\
 & \;{\textstyle \sum_{i\in\mathcal{N}}}\mathbf{v}_{ik}\herm\mathbf{v}_{ik}\leq P_{k},\forall k\in\mathcal{K}\label{eq:powerconst-1}
\end{align}
\end{subequations} where $\{\mu_{i}\geq0\}$ are newly introduced
variables.
\begin{lem}
\label{lem:equivalentIA}Let $(\{\mathbf{v}_{ik}^{\star}\},\{\mu_{i}^{\star}\})$
be an optimal solution to (\ref{eq:EE_prob-1}), then $\{\mathbf{v}_{ik}^{\star}\}$
is an optimal solution to (\ref{eq:wsr_prob}). Conversely, let $\{\mathbf{v}_{ik}^{\ast}\}$
be an optimal solution to (\ref{eq:wsr_prob}), then $(\{\mathbf{v}_{ik}^{\ast}\},\{\gamma_{i}(\{\mathbf{v}_{ik}^{\ast}\})\})$
is an optimal solution to (\ref{eq:EE_prob-1}). Moreover, the two
problems have the same optimal objective value.
\end{lem}
\begin{IEEEproof}
A proof is provided in Appendix \ref{subsec:Proof-of-eqiIA}.
\end{IEEEproof}
It is clear that the nonconvexity of \eqref{eq:wsr_prob-1} is due
to (\ref{eq:snrconstr}), which can be equivalently rewritten as 
\begin{equation}
\eqref{eq:snrconstr}\Leftrightarrow\begin{cases}
{\textstyle \sum_{k\in\mathcal{K}}}|\mathbf{h}_{ik}\mathbf{v}_{ik}|^{2}/u_{i}\geq\mu_{i}\\
{\textstyle \sum_{k\in\mathcal{K}}\sum_{\substack{j\in\mathcal{N}\setminus\{i\}}
}}|\mathbf{h}_{ik}\mathbf{v}_{jk}|^{2}+\sigma_{i}^{2}\leq u_{i}
\end{cases}
\end{equation}
where $\{u_{i}>0\}$ are slack variables. Note that the quadratic-over-linear
function is convex with the involved variables. We introduce $\{u_{i}\}$
to avoid the function $\sum_{k\in\mathcal{K}}|\mathbf{h}_{ik}\mathbf{v}_{ik}|^{2}/\mu_{i}$,
which might lead to numerical problems, since $\mu_{i}$ could be
zero. In light of the InAp approach, we use a first order approximation
as a convex lower bound to derive an approximate convex problem. More
explicitly, let $(\{\mathbf{v}_{i}^{(t)}\},\{\mu_{i}^{(t)}\},\{u_{i}^{(t)}\})$
be a feasible point, then the approximate problem is \begin{subequations}\label{eq:EE_prob-1-2}
\begin{align}
\underset{\{\mathbf{v}_{ik}\},\{\mu_{i}\},\{u_{i}\}}{\maxi} & \;{\textstyle \sum_{i\in\mathcal{N}}}w_{i}\log(1+\mu_{i})\label{eq:GeneralObj-1-2}\\
\st & \;{\textstyle \sum_{k\in\mathcal{K}}}(\Re\{\mathbf{g}_{ik}^{(t)}\mathbf{v}_{ik}\}-A_{ik}^{(t)}u_{i})\geq\mu_{i},\forall i\in\mathcal{N},\label{eq:snrconstr-2}\\
 & \;{\displaystyle {\textstyle \sum_{k\in\mathcal{K}}\sum_{\substack{j\in\mathcal{N}\setminus\{i\}}
}}}|\mathbf{h}_{ik}\mathbf{v}_{jk}|^{2}+\sigma_{i}^{2}\leq u_{i},\forall i\in\mathcal{N},\label{eq:interf}\\
 & \;{\textstyle \sum_{i\in\mathcal{N}}}\mathbf{v}_{ik}\herm\mathbf{v}_{ik}\leq P_{k},\forall k\in\mathcal{K}\label{eq:powerconst-1-2}
\end{align}
\end{subequations} where $\mathbf{g}_{ik}^{(t)}=(2/u_{i}^{(t)})(\mathbf{v}_{ik}^{(t)})\herm\mathbf{h}_{ik}\herm\mathbf{h}_{ik}$
and $A_{ik}^{(t)}=(|\mathbf{h}_{ik}\mathbf{v}_{ik}^{(t)}|/u_{i}^{(t)})^{2}$.

\subsubsection*{CQP-Based Approximation}

Since $w_{i}$, $i\in\mathcal{N}$, is generally different, problem
\eqref{eq:EE_prob-1-2} containing a mix of exponential and second-order
cones is normally treated as a generic convex program. The efficiency
of modern convex solvers in solving these generic programs is far
less than in solving more standard ones. This motivates us to present
a quadratic approximation of the objective to obtain a CQP approximate
problem of \eqref{eq:EE_prob-1-2}. We achieve this by using a lower
bound of the logarithm function given as 
\begin{equation}
\log(1+\mu_{i})\geq\log(1+\mu_{i}^{(t)})+2-2\sqrt{(1+\mu_{i}^{(t)})/(1+\mu_{i})}.\label{eq:log_appx}
\end{equation}
The validity of the bound according to the InAp principles is justified
in our recent work \cite[Sec. III-E]{giangjsac}. Next, by introducing
new variables $\{\delta_{i}\}$ and $\{\pi_{i}\}$, we arrive at the
following CQP approximation \begin{subequations}\label{eq:EE_prob-1-2-1}
\begin{align}
\underset{\mathbf{x}}{\mini} & \;{\textstyle \sum_{i\in\mathcal{N}}}\tilde{w}_{i}^{(t)}\pi_{i}\label{eq:GeneralObj-1-2-1}\\
\st & \;\parallel[2,(\pi_{i}-\delta_{i})]\parallel_{2}\leq(\pi_{i}+\delta_{i}),1+\mu_{i}\geq\delta_{i}^{2},\delta_{i}\geq1,\forall i\in\mathcal{N},\label{eq:log_appr}\\
 & \;\eqref{eq:snrconstr-2},\eqref{eq:interf},\eqref{eq:powerconst-1-2}\label{eq:snrconstr-2-1}
\end{align}
\end{subequations}where $\tilde{w}_{i}^{(t)}=w_{i}\sqrt{1+\mu_{i}^{(t)}}$,
$\mathbf{x}\triangleq\{\{\mathbf{v}_{ik}\},\{\mu_{i}\},\{u_{i}\},\{\delta_{i}\},\{\pi_{i}\}\}$.

\subsubsection*{Algorithm and Convergence}

The InAp-based iterative procedure is outlined in Algorithm \ref{alg:IA_alg},
which starts with a random initial point (Step \ref{alg:initial}).
In each iteration, a CQP is solved (Step \ref{alg:solvesca}) and
the feasible point is updated (Step \ref{alg:updatesca}). Successively
solving \eqref{eq:EE_prob-1-2-1} and updating $(\{\mathbf{v}_{i}^{(t)}\},\{\mu_{i}^{(t)}\},\{u_{i}^{(t)}\})$
by the optimal solution of \eqref{eq:EE_prob-1-2-1}, we obtain the
objective sequence $\{\sum_{i\in\mathcal{N}}w_{i}\log(1+\mu_{i}^{(t)})\}_{t=0}^{\infty}$
which is guaranteed to converge as stated in the following lemma.
\begin{lem}
\label{lem:ConvergeIA}Any sequence $\{\sum_{i\in\mathcal{N}}w_{i}\log(1+\mu_{i}^{(t)})\}_{t=0}^{\infty}$
produced by Algorithm \ref{alg:IA_alg} is monotonically increasing
and converges.
\end{lem}
\begin{IEEEproof}
A proof is provided in Appendix \ref{subsec:Proof-ConvergeIA}.
\end{IEEEproof}
\begin{algorithm}[!t]
\caption{An InAp Algorithm to Efficiently Solve (\ref{eq:wsr_prob}\foreignlanguage{english}{)}}
\label{alg:IA_alg} \begin{algorithmic}[1]

\STATE \textbf{Initialization}: Set small $\epsilon_{\mathrm{IA}}$,
set $t\coloneqq0$, choose initial $(\{\mathbf{v}_{i}^{(0)}\},\{\mu_{i}^{(0)}\},\{u_{i}^{(0)}\})$.\label{alg:initial}

\REPEAT 

\STATE Solve \eqref{eq:EE_prob-1-2-1}, and denote the optimal solution
by $(\{\mathbf{v}_{i}^{\ast}\},\{\mu_{i}^{\ast}\},\{u_{i}^{\ast}\})$\label{alg:solvesca}

\STATE Update $(\{\mathbf{v}_{i}^{(t+1)}\},\{\mu_{i}^{(t+1)}\},\{u_{i}^{(t+1)}\}):=(\{\mathbf{v}_{i}^{\ast}\},\{\mu_{i}^{\ast}\},\{u_{i}^{\ast}\})$\label{alg:updatesca}

\STATE Update $t:=t+1$

\UNTIL Convergence on objective value 

\STATE \textbf{Output}: $\{\mathbf{v}_{i}^{(t)}\}$

\end{algorithmic}
\end{algorithm}

\subsubsection*{Computational Complexity}

The computational complexity of the algorithm depends on the arithmetical
cost of solving approximate problem \eqref{eq:EE_prob-1-2-1} in each
iteration. Problem \eqref{eq:EE_prob-1-2-1} includes $2N(2+M)$ real
variables where $M=\sum_{k\in\mathcal{K}}M_{k}$, $2N$ constraints
in size 3, $2N$ constraints in size 1, $N$ constraints in size $K(N-1)+N+2$
, and one constraint in size $2M_{k}+1$ for BS $k$, $\forall k\in\mathcal{K}$.
Hence, the worst-case computational cost of using a general interior
point method for solving \eqref{eq:EE_prob-1-2-1} is $\mathcal{O}(\sqrt{1+5N+K}4N^{2}(2+M)^{2}(N^{2}(K+1)+N(10-K)+2M+K))$
\cite{SOCApp_boyd1999}.
\begin{rem}
\label{rem:A-first-order-solution}\emph{{[}A first-order solution
to \eqref{eq:wsr_prob}{]}} We have introduced a few slack variables
to achieve a CQP approximation. This maneuver certainly increases
the complexity of the problem and may question the efficacy of the
proposed iterative solution. This concern is especially relevant as
the feasible set of \eqref{eq:wsr_prob} is expressed as a system
of \emph{separable} quadratic convex constraints. Thus it is apparently
appealing to approximate the objective of \eqref{eq:wsr_prob} by
means of a first-order optimization method. In this way, the resulting
program in each iteration has low complexity compared to \eqref{eq:EE_prob-1-2-1}.
One of the first-order methods widely used for a nonconvex problem
such as \eqref{eq:wsr_prob} is the conditional gradient technique
(a.k.a.\ the Frank-Wolfe method) that have received significant interest
recently \cite{convergenonconvexFW,Ho_DFW_2018,adaptstepsizeFW}.
In Appendix \ref{sec:A-First-Order-Algorithm} we show how a Frank-Wolfe
(FW) type algorithm can be derived to solve \eqref{eq:wsr_prob},
where the linear optimization oracle at each iteration admits a closed-form
expression. While looking very attractive from a per-iteration cost
viewpoint, FW type methods in general converge very slowly, i.e.,
they need a very large number of iterations to produce a high-accuracy
solution. As a consequence, the actual run time of the FW type algorithm
is much higher than our proposed solution presented in Algorithm \ref{alg:IA_alg}.
We provide numerical examples to demonstrate this point in Fig.\ \ref{fig:convergeFW}.
\end{rem}

\section{Distributed Implementation: A Combination of InAp and ADMM \label{sec:Distributed-Algorithm-via}}

In this section, we develop a decentralized algorithm implementing
the InAp-based solution where beamforming vectors are calculated locally
at the MEC servers using local CSI. The approach is to use the ADMM
to solve the convex approximation subproblem \eqref{eq:EE_prob-1-2-1},
in which \eqref{eq:EE_prob-1-2-1} is converted to an equivalent transformation
so that the ADMM procedure can be distributively implemented.

\subsection{Distributed Formulation}

We first rearrange \eqref{eq:EE_prob-1-2-1} according to the MEC
servers as \begin{subequations}\label{eq:EE_prob-1-2-1-1} 
\begin{align}
\underset{\mathbf{x}}{\mini} & \;{\textstyle \sum_{i\in\mathcal{N}}}\tilde{w}_{i}^{(t)}\pi_{i}\label{eq:GeneralObj-1-2-1-1}\\
\st & \;\parallel[2,(\pi_{i}-\delta_{i})]\parallel_{2}\leq(\pi_{i}+\delta_{i}),1+\mu_{i}\geq\delta_{i}^{2},\delta_{i}\geq1,\forall i\in\mathcal{N},\label{eq:log_appr-1}\\
 & \;{\textstyle \sum_{d\in\mathcal{D}}\sum_{k\in\mathcal{K}_{d}}}(\Re\{\mathbf{g}_{ik}^{(t)}\mathbf{v}_{ik}\}-A_{ik}^{(t)}u_{i})\geq\mu_{i},\forall i\in\mathcal{N},\label{eq:constrcsig}\\
 & \;{\textstyle \sum_{d\in\mathcal{D}}\sum_{k\in\mathcal{K}_{d}}\sum_{\substack{j\in\mathcal{N}\setminus\{i\}}
}}|\mathbf{h}_{ik}\mathbf{v}_{jk}|^{2}+\sigma_{i}^{2}\leq u_{i},\forall i\in\mathcal{N},\label{eq:intersig}\\
 & \;{\textstyle \sum_{i\in\mathcal{N}}}\mathbf{v}_{ik}\herm\mathbf{v}_{ik}\leq P_{k},\forall k\in\mathcal{K}.\label{eq:powerref}
\end{align}
\end{subequations}We observe that \eqref{eq:constrcsig} and \eqref{eq:intersig}
are the coupling constraints. Thus, we introduce local and global
variables so that these constraints are decoupled among the MEC servers.
In particular, we equivalently rewrite \eqref{eq:EE_prob-1-2-1-1}
into the following form \begin{subequations}\label{eq:EE_prob-1-1-1}
\begin{align}
\underset{\substack{\{\mathbf{v}_{ik}\},\{\mu_{i}\},\{u_{i}\},\{\delta_{i}\},\\
\{\pi_{i}\},\{\hat{q}_{ik}\},\{\tilde{q}_{ik}\},\{q_{ik}\},\\
\{\hat{y}_{ik}\},\{\tilde{y}_{ik}\},\{y_{ik}\}
}
}{\mini} & \;{\textstyle \sum_{i\in\mathcal{N}}}\tilde{w}_{i}^{(t)}\pi_{i}\label{eq:GeneralObj-1-1-1}\\
\st & \;{\textstyle \sum_{k\in\mathcal{K}_{1}}}\Re\{\mathbf{g}_{ik}^{(t)}\mathbf{v}_{ik}\}-\left({\textstyle \sum_{k\in\mathcal{K}_{1}}}A_{ik}^{(t)}+{\textstyle \sum_{d\in\bar{\mathcal{D}}}\sum_{k\in\mathcal{K}_{d}}}A_{ik}^{(t)}\right)u_{i}\nonumber \\
 & \;\hspace{7cm}+{\textstyle \sum_{d\in\bar{\mathcal{D}}}}\tilde{y}_{id}\geq\mu_{i},\forall i\in\mathcal{N},\label{eq:snrconstr-1-1}\\
 & \;{\textstyle \sum_{k\in\mathcal{K}_{d}}}\Re\{\mathbf{g}_{ik}^{(t)}\mathbf{v}_{ik}\}\geq\hat{y}_{id},\forall i\in\mathcal{N},d\in\bar{\mathcal{D}}\label{eq:yconstr}\\
 & \;{\textstyle \sum_{k\in\mathcal{K}_{1}}\sum_{\substack{j\in\mathcal{N}\setminus\{i\}}
}}|\mathbf{h}_{ik}\mathbf{v}_{jk}|^{2}+{\textstyle \sum_{d\in\bar{\mathcal{D}}}}\tilde{q}_{id}+\sigma_{i}^{2}\leq u_{i},\forall i\in\mathcal{N},\label{eq:snrcontr2}\\
 & \;{\textstyle \sum_{k\in\mathcal{K}_{d}}\sum_{\substack{j\in\mathcal{N}\setminus\{i\}}
}}|\mathbf{h}_{ik}\mathbf{v}_{jk}|^{2}\leq\hat{q}_{id},\forall i\in\mathcal{N},d\in\bar{\mathcal{D}}\label{eq:qconstr}\\
 & \;\hat{q}_{id}=q_{id},\hat{y}_{id}=y_{id},\forall i\in\mathcal{N},\forall d\in\bar{\mathcal{D}}\label{eq:equalconstr1}\\
 & \;\tilde{q}_{id}=q_{id},\tilde{y}_{ik}=y_{id},\forall i\in\mathcal{N},\forall d\in\bar{\mathcal{D}}\label{eq:equalconstr2}\\
 & \;\eqref{eq:powerref},\eqref{eq:log_appr-1}
\end{align}
\end{subequations} where $\bar{\mathcal{D}}=\mathcal{D}\setminus\{1\}$,
and $\{\hat{q}_{id}\}$, $\{\tilde{q}_{id}\}$, $\{q_{id}\}$, $\{\hat{y}_{id}\}$,
$\{\tilde{y}_{id}\}$ and $\{y_{id}\}$ are newly introduced variables
for decomposing \eqref{eq:constrcsig} and \eqref{eq:intersig} into
constraints which will be handled locally at the MEC servers; constraints
\eqref{eq:equalconstr1} and \eqref{eq:equalconstr2} ensure the agreement
of the local variables $\{\hat{q}_{id}\}$ and $\{\tilde{q}_{id}\}$,
and $\{\hat{y}_{id}\}$ and $\{\tilde{y}_{id}\}$.
\begin{lem}
\label{lem:equiADMM}Let $(\mathbf{x}^{\star},\{\hat{q}_{id}^{\star}\},\{\tilde{q}_{id}^{\star}\},\{q_{id}^{\star}\},\{\hat{y}_{id}^{\star}\},\{\tilde{y}_{id}^{\star}\},\{y_{id}^{\star}\})$
be an optimal solution to \eqref{eq:EE_prob-1-1-1}, then $\mathbf{x}^{\star}$
is an optimal solution to \eqref{eq:EE_prob-1-2-1-1}. Conversely,
let $\mathbf{x}^{\ast}$ be an optimal solution to \eqref{eq:EE_prob-1-2-1-1},
then $(\mathbf{x}^{\ast},\{\hat{q}_{id}^{\ast}\},\{\tilde{q}_{id}^{\ast}\},\{q_{id}^{\ast}\},\{\hat{y}_{id}^{\ast}\},\{\tilde{y}_{id}^{\ast}\},\{y_{id}^{\ast}\})$
where $\tilde{q}_{id}^{\ast}=q_{id}^{\ast}=\hat{q}_{id}^{\ast}={\textstyle \sum_{k\in\mathcal{K}_{d}}\sum_{\substack{j\in\mathcal{N}\setminus\{i\}}
}}|\mathbf{h}_{ik}\mathbf{v}_{jk}^{\ast}|^{2}$ and $\tilde{y}_{id}^{\ast}=y_{id}^{\ast}=\hat{y}_{id}^{\ast}={\textstyle \sum_{k\in\mathcal{K}_{d}}}\Re\{\mathbf{g}_{ik}^{(t)}\mathbf{v}_{ik}^{\ast}\}$,
$\forall i\in\mathcal{N},d\in\bar{\mathcal{D}}$, is an optimal solution
to \eqref{eq:EE_prob-1-1-1}.
\end{lem}
\begin{IEEEproof}
The lemma can be proved by using the similar approach to that of Lemma
\ref{lem:equivalentIA}. The detail is omitted for the sake of brevity.
\end{IEEEproof}
We now rewrite \eqref{eq:EE_prob-1-1-1} in a more compact form. Without
loss of generality, we assume that macro BS is controlled by MEC server
1 and $\bar{\mathcal{D}}=\mathcal{D}\setminus\{1\}$. For notational
convenience, let us denote by $\tilde{\mathbf{x}}\triangleq\{\{\mathbf{v}_{ik}\}_{i\in\mathcal{N},k\in\mathcal{K}_{1}},\{\mu_{i}\}_{i\in\mathcal{N}},\{u_{i}\}_{i\in\mathcal{N}},\{\delta_{i}\}_{i\in\mathcal{N}},\{\pi_{i}\}_{i\in\mathcal{N}},\{\tilde{q}_{id}\}_{i\in\mathcal{N},d\in\bar{\mathcal{D}}},\{\tilde{y}_{id}\}_{i\in\mathcal{N},d\in\bar{\mathcal{D}}}\}$
the local variables at MEC server 1, and define its local feasible
set as 
\begin{align}
\tilde{\mathcal{S}} & \triangleq\{\tilde{\mathbf{x}}|\eqref{eq:snrconstr-1-1},\eqref{eq:snrcontr2},\eqref{eq:log_appr},{\textstyle \sum_{i\in\mathcal{N}}}\mathbf{v}_{ik}\herm\mathbf{v}_{ik}\leq P_{k},\forall k\in\mathcal{K}_{1}\}.\label{eq:setbs-1}
\end{align}
Similarly, let us denote by $\hat{\mathbf{x}}_{d}\triangleq\{\{\mathbf{v}_{ik}\}_{i\in\mathcal{N},k\in\mathcal{K}_{d}},\{\hat{q}_{id}\}_{i\in\mathcal{N}},\{\hat{y}_{id}\}_{i\in\mathcal{N}}\}$
the local variables at MEC server $d$, $d\in\bar{\mathcal{D}}$,
and define its local feasible set as 
\begin{multline}
\hat{\mathcal{S}}_{d}=\{\hat{\mathbf{x}}_{d}|{\textstyle \sum_{k\in\mathcal{K}_{d}}}\Re\{\mathbf{g}_{ik}^{(t)}\mathbf{v}_{ik}\}\geq\hat{y}_{id},\forall i\in\mathcal{N},{\textstyle \sum_{k\in\mathcal{K}_{d}}\sum_{\substack{j\in\mathcal{N}\setminus\{i\}}
}}|\mathbf{h}_{ik}\mathbf{v}_{jk}|^{2}\leq\hat{q}_{id},\forall i\in\mathcal{N},\\
{\textstyle \sum_{i\in\mathcal{N}}}\mathbf{v}_{ik}\herm\mathbf{v}_{ik}\leq P_{k},\forall k\in\mathcal{K}_{d}\}.\label{eq:setbs}
\end{multline}
With these notations, we can rewrite \eqref{eq:EE_prob-1-1-1} as
\begin{subequations}\label{eq:EE_prob-1-1-1-1} 
\begin{align}
\underset{\substack{\{\tilde{\mathbf{x}}\in\tilde{\mathcal{S}}\},\{\hat{\mathbf{x}}_{d}\in\hat{\mathcal{S}}_{d}\}_{d\in\bar{\mathcal{D}}}\\
\{q_{id}\}_{i\in\mathcal{N},d\in\bar{\mathcal{D}}},\{y_{id}\}_{i\in\mathcal{N},d\in\bar{\mathcal{D}}}
}
}{\mini} & \;{\textstyle \sum_{i\in\mathcal{N}}}\tilde{w}_{i}^{(t)}\pi_{i}\label{eq:GeneralObj-1-1-1-1}\\
\st & \;\tilde{\boldsymbol{\pi}}=\tilde{\boldsymbol{\phi}},\hat{\boldsymbol{\pi}}_{d}=\hat{\boldsymbol{\phi}}_{d},\forall d\in\bar{\mathcal{D}}
\end{align}
\end{subequations} where $\tilde{\boldsymbol{\pi}}\triangleq\{\{\tilde{q}_{id}\}_{i\in\mathcal{N},d\in\bar{\mathcal{D}}},\{\tilde{y}_{id}\}_{i\in\mathcal{N},d\in\bar{\mathcal{D}}}\}$,
$\hat{\boldsymbol{\pi}}_{d}\triangleq\{\{\hat{q}_{id}\}_{i\in\mathcal{N}},\{\tilde{y}_{id}\}_{i\in\mathcal{N}}\}$;
$\tilde{\boldsymbol{\phi}}$ and $\hat{\boldsymbol{\phi}}_{d}$ are
the rearranged vectors from the same set of global variables $(\{q_{id}\}_{i\in\mathcal{N},d\in\bar{\mathcal{D}}},\{y_{id}\}_{i\in\mathcal{N},d\in\bar{\mathcal{D}}})$.

Now, it can be seen that (\ref{eq:EE_prob-1-1-1-1}) is in the form
of consensus problem which can be solved using the ADMM \cite{ADMM_boyd}.
We have the augmented Lagrangian function of (\ref{eq:EE_prob-1-1-1-1})
given by
\begin{multline}
\mathfrak{L}(\{\tilde{\mathbf{x}}\},\{\hat{\mathbf{x}}_{d}\},\{q_{id}\},\{y_{id}\};\{\boldsymbol{\xi}\},\{\boldsymbol{\rho}_{d}\})=\Bigl(\boldsymbol{\xi}\trans(\tilde{\boldsymbol{\pi}}-\tilde{\boldsymbol{\phi}})+\frac{m}{2}||\tilde{\boldsymbol{\pi}}-\tilde{\boldsymbol{\phi}}||_{2}^{2}+{\textstyle \sum_{i\in\mathcal{N}}}\tilde{w}_{i}^{(t)}\pi_{i}\Bigr)\\
+{\textstyle \sum_{d\in\bar{\mathcal{D}}}}\left(\boldsymbol{\rho}_{d}\trans(\hat{\boldsymbol{\pi}}_{d}-\hat{\boldsymbol{\phi}}_{d})+\frac{m}{2}||\hat{\boldsymbol{\pi}}_{d}-\hat{\boldsymbol{\phi}}_{d}||_{2}^{2}\right)\label{eq:lagrangian-small}
\end{multline}
where $\boldsymbol{\xi}\in\mathbb{R}^{N(D-1)\times1}$ and $\{\boldsymbol{\rho}_{d}\}_{d\in\bar{\mathcal{D}}}$,$\boldsymbol{\rho}_{d}\in\mathbb{R}^{N\times1}$,
are the vectors of Lagrangian multipliers; $m>0$ is the penalty parameter
that weighs the violation of the equality constraints. In what follows,
we present the variable updates at iteration $(j+1)$ of the ADMM
procedure.

\subsection{Variable Updates}

\subsubsection{Update Local Variables}

Let $\boldsymbol{\xi}^{(j)}$, $\tilde{\boldsymbol{\phi}}^{(j)}$,
and $\{\boldsymbol{\rho}_{d}^{(j)}\}$ be the values obtained at iteration
$j$. MEC server 1 updates its local variables $\tilde{\mathbf{x}}$
by solving the following CQP
\begin{align}
\underset{\tilde{\mathbf{x}}\in\tilde{\mathcal{S}}}{\mini}\;(\boldsymbol{\xi}^{(j)})\trans(\tilde{\boldsymbol{\pi}}-\tilde{\boldsymbol{\phi}}^{(j)})+\frac{m}{2}||\tilde{\boldsymbol{\pi}}-\tilde{\boldsymbol{\phi}}^{(j)}||_{2}^{2}+{\textstyle \sum_{i\in\mathcal{N}}}\tilde{w}_{i}^{(t)}\pi_{i}.\label{eq:macupdate-small}
\end{align}
Also, MEC server $d$, $d\in\bar{\mathcal{D}}$, updates its local
variables $\hat{\mathbf{x}}_{d}$ by solving the following quadratically
constrained quadratic program (QCQP)
\begin{align}
\underset{\hat{\mathbf{x}}_{d}\in\hat{\mathcal{S}}_{d}}{\mini}\;(\boldsymbol{\rho}_{d}^{(j)})\trans(\hat{\boldsymbol{\pi}}_{d}-\hat{\boldsymbol{\phi}}_{d}^{(j)})+\frac{m}{2}||\hat{\boldsymbol{\pi}}_{d}-\hat{\boldsymbol{\phi}}_{d}^{(j)}||_{2}^{2}.\label{eq:bsupdate-small}
\end{align}

\subsubsection{Update Global Variables}

The global variables $\{q_{id}\}$ and $\{y_{id}\}$ are updated via
finding the minimum of the following quadratic function derived from
\eqref{eq:lagrangian-small} 
\begin{multline}
G^{(j)}(\{q_{id}\},\{y_{id}\})\triangleq{\textstyle \sum_{d\in\bar{\mathcal{D}}}\sum_{i\in\mathcal{N}}}\left([\boldsymbol{\xi}^{(j)}]_{q_{id}}(\tilde{q}_{id}^{(j+1)}-q_{id})+\frac{m}{2}(\tilde{q}_{id}^{(j+1)}-q_{id})^{2}\right.\\
+[\boldsymbol{\xi}^{(j)}]_{y_{id}}(\tilde{y}_{id}^{(j+1)}-y_{id})+\frac{m}{2}(\tilde{y}_{id}^{(j+1)}-y_{id})^{2}+[\boldsymbol{\rho}_{d}^{(j)}]_{q_{id}}(\hat{q}_{id}^{(j+1)}-q_{id})+\frac{m}{2}(\hat{q}_{id}^{(j+1)}-q_{id})^{2}\\
\left.+[\boldsymbol{\rho}_{d}^{(j)}]_{y_{id}}(\hat{y}_{id}^{(j+1)}-y_{id})+\frac{m}{2}(\hat{y}_{id}^{(j+1)}-y_{id})^{2}\right)\label{eq:globalupdate-small}
\end{multline}
where $[\boldsymbol{\xi}^{(j)}]_{q_{id}}$ is the element in $\boldsymbol{\xi}^{(j)}$
corresponding to constraint $\tilde{q}_{id}=q_{id}$; similar definition
is applied to $[\boldsymbol{\xi}^{(j)}]_{y_{id}}$, $[\boldsymbol{\rho}_{k}^{(j)}]_{q_{id}}$
and $[\boldsymbol{\rho}_{d}^{(j)}]_{y_{id}}$. The closed-form of
the minimizer of \eqref{eq:globalupdate-small} is given as

\begin{align}
q_{id}^{(j+1)} & =\frac{([\boldsymbol{\xi}^{(j)}]_{q_{id}}+m\tilde{q}_{id}^{(j+1)})+([\boldsymbol{\rho}_{d}^{(j)}]_{q_{id}}+m\hat{q}_{id}^{(j+1)})}{2m}\label{eq:update-s}\\
y_{id}^{(j+1)} & =\frac{([\boldsymbol{\xi}^{(j)}]_{y_{id}}+m\tilde{y}_{id}^{(j+1)})+([\boldsymbol{\rho}_{d}^{(j)}]_{y_{id}}+m\hat{y}_{id}^{(j+1)})}{2m}.\label{eq:update-z-small}
\end{align}

\subsubsection{Update Lagrangian Multipliers}

The Lagrangian multipliers are updated as follows 
\begin{align}
\boldsymbol{\xi}^{(j+1)} & =\boldsymbol{\xi}^{(j)}+m(\tilde{\boldsymbol{\pi}}^{(j+1)}-\tilde{\boldsymbol{\phi}}^{(j+1)})\label{eq:xiupdate-small}\\
\boldsymbol{\rho}_{d}^{(j+1)} & =\boldsymbol{\rho}_{d}^{(j)}+m(\hat{\boldsymbol{\pi}}_{d}^{(j+1)}-\hat{\boldsymbol{\phi}}_{d}^{(j+1)})\label{eq:rhoupdate-small}
\end{align}
where $\boldsymbol{\xi}^{(j+1)}$ can be determined at MEC server
1 while $\boldsymbol{\rho}_{d}^{(j+1)}$ is determined at MEC $d$,
$d\in\bar{\mathcal{D}}$.

\subsection{The Distributed Algorithm}

We summarize the proposed distributed algorithm in Algorithm \ref{alg:Dist_impl-small}.
It includes two stages: the inner stage is the ADMM procedure solving
InAp subproblems, and the outer stage is the InAp feasible point update
using the values obtained at the inner stage (Step \ref{alg:updatesca}).
The values obtained in the last iteration of the ADMM at InAp iteration
$t$ are used for initializing ADMM procedure at InAp iteration $t+1$
(Step \ref{alg:updateini}). The initial values for the algorithm
(Step \ref{alg:initial}) will be specified in Section \ref{sec:Numerical-Results}.
\begin{algorithm}[!t]
\caption{Decentralized Procedure Solving (\ref{eq:wsr_prob}\foreignlanguage{english}{)}}
\label{alg:Dist_impl-small} \begin{algorithmic}[1]

\STATE \textbf{Initialization}: Set small $\epsilon_{\mathrm{IA}}$,
$t\coloneqq0$ and $j\coloneqq0$, choose initial values for $(\{\mathbf{v}_{ik}^{(0)}\},\{\mu_{i}^{(0)}\},\{u_{i}^{(0)}\})$
and $(\{q_{id}^{(0)}\},\{y_{id}^{(0)}\};\boldsymbol{\xi}^{(0)},\{\boldsymbol{\rho}_{d}^{(0)}\})$.\label{alg:initial}

\REPEAT[Outer stage (SCA procedure)] 

\STATE MEC server $d$, $d\in\bar{\mathcal{D}}$, receives $\{u_{i}^{(t)}\}$
from MEC server 1 to form $\sum_{k\in\mathcal{K}_{d}}A_{ik}^{(t)}$
and $\mathbf{g}_{ik}^{(t)}$.

\STATE MEC server 1 receives scalar $\sum_{k\in\mathcal{K}_{d}}A_{ik}^{(t)}$
from MEC server $d$, $d\in\bar{\mathcal{D}}$, to form $\sum_{d\in\bar{\mathcal{D}}}\sum_{k\in\mathcal{K}_{d}}A_{ik}^{(t)}$.

\REPEAT[Inner stage (ADMM procedure)] 

\STATE MEC server 1 updates $\tilde{\mathbf{x}}^{(j+1)}$ by solving
\eqref{eq:macupdate-small}; MEC server d, $d\in\bar{\mathcal{D}}$,
updates $\hat{\mathbf{x}}_{d}^{(j+1)}$ by solving \eqref{eq:bsupdate-small}.

\STATE MEC server 1 receives scalars $\frac{([\boldsymbol{\rho}_{d}^{(j)}]_{q_{id}}+m\hat{q}_{id}^{(j+1)})}{2m}$
and $\frac{\left([\boldsymbol{\rho}_{d}^{(j)}]_{y_{id}}+m\hat{y}_{id}^{(j+1)}\right)}{2m}$
from MEC server $d$, $d\in\bar{\mathcal{D}}$, then updates global
variables $q_{id}^{(j+1)}$ and $y_{id}^{(j+1)}$ using \eqref{eq:update-s}
and \eqref{eq:update-z-small}.

\STATE MEC server 1 updates $\boldsymbol{\xi}^{(j+1)}$ by \eqref{eq:xiupdate-small};
MEC server $d$, $d\in\bar{\mathcal{D}}$, receives $q_{id}^{(j+1)}$
and $y_{id}^{(j+1)}$ from MEC server 1, then updates $\boldsymbol{\rho}_{d}^{(j+1)}$
using \eqref{eq:rhoupdate-small}.\label{alg:updatemultipliers}

\STATE $j:=j+1$.

\UNTIL {ADMM convergence}

\STATE Obtain $(\{\mathbf{v}_{ik}^{\ast}\},\{\mu_{i}^{\ast}\},\{u_{i}^{\ast}\},\{q_{id}^{\ast}\},\{y_{id}^{\ast}\};\boldsymbol{\xi}^{\ast},\{\boldsymbol{\rho}_{d}^{\ast}\})$,
the solution from the ADMM procedure. 

\STATE Update $t:=t+1$, $j\coloneqq0$

\STATE Update $(\{\mathbf{v}_{ik}^{(t)}\},\{\mu_{i}^{(t)}\},\{u_{i}^{(t)}\}):=(\{\mathbf{v}_{ik}^{\ast}\},\{\mu_{i}^{\ast}\},\{u_{i}^{\ast}\})$\label{alg:updatesca}

\STATE Update $(\{q_{id}^{(0)}\},\{y_{id}^{(0)}\};\boldsymbol{\xi}^{(0)},\{\boldsymbol{\rho}_{d}^{(0)}\})\coloneqq(\{q_{id}^{\ast}\},\{y_{id}^{\ast}\};\boldsymbol{\xi}^{\ast},\{\boldsymbol{\rho}_{d}^{\ast}\})$\label{alg:updateini}

\UNTIL {$\sum_{i\in\mathcal{N}}w_{i}(\log(1+\mu_{i}^{(t+1)})-\log(1+\mu_{i}^{(t)}))\leq\epsilon_{\mathrm{IA}}$}
\end{algorithmic} 
\end{algorithm}

\subsubsection{Exchanged Signals}

We now discuss the signaling exchanged between MEC servers for implementing
Algorithm \ref{alg:Dist_impl-small}. In each ADMM iteration (inner
stage), MEC server 1 acquires the two scalars $\frac{[\boldsymbol{\rho}_{d}^{(j)}]_{q_{id}}+m\hat{q}_{id}^{(j+1)}}{2m}$
and $\frac{[\boldsymbol{\rho}_{d}^{(j)}]_{y_{id}}+m\hat{y}_{id}^{(j+1)}}{2m}$
from MEC server $d$, $d\in\bar{\mathcal{D}}$, to updates the global
variables $q_{id}^{(j+1)}$ and $y_{id}^{(j+1)}$ . In order to update
multipliers $\boldsymbol{\rho}_{d}^{(j+1)}$, $d\in\bar{\mathcal{D}}$,
MEC server $d$ needs the global parameters $q_{id}^{(j+1)}$ and
$y_{id}^{(j+1)}$ from MEC server 1. For the outer stage (InAp iteration),
after the inner stage converges, MEC server $d$, $d\in\bar{\mathcal{D}}$,
needs $\{u_{i}^{(t)}\}$ from MEC server 1 to update $\sum_{k\in\mathcal{K}_{d}}A_{ik}^{(t)}$
and $\mathbf{g}_{ik}^{(t)}$; and MEC server 1 needs scalar $\sum_{k\in\mathcal{K}_{d}}A_{ik}^{(t)}$
from MEC server $d$, $d\in\bar{\mathcal{D}}$. From these, the exchanged
signal overhead depends on the numbers of MEC servers and users, and
is independent from the number of BSs or the number of transmit antenna.

\subsubsection{Convergence of Algorithm \ref{alg:Dist_impl-small}}

The convergence of Algorithm \ref{alg:Dist_impl-small} depends on
that of the outer and inner stages. As discussed in Section \ref{sec:Efficient-Solutions},
the outer stage procedure converges when the convex approximate problems
are solved optimally. This is achieved by the inner stage procedure
as stated in the following lemma.
\begin{lem}
\label{lem:convergeADMM}The ADMM procedure of Algorithm \ref{alg:Dist_impl-small}
guarantees to output a solution which achieves the optimal objective
value of \eqref{eq:EE_prob-1-1-1}.
\end{lem}
\begin{IEEEproof}
A proof is provided in Appendix \ref{sec:Proof-of-ConvergeADMM}.
\end{IEEEproof}
Commonly, the convergence of the ADMM procedure is observed via the
primal and dual residuals \cite[Section 3.3.1]{ADMM_boyd}. Specifically,
let us define local primal and dual residual vectors as
\[
\boldsymbol{\varepsilon}_{d}^{\mathrm{pri}}\triangleq\left\{ \begin{array}{ll}
\tilde{\boldsymbol{\pi}}^{(j)}-\tilde{\boldsymbol{\phi}}^{(j)} & \textrm{for }d=1\\
\hat{\boldsymbol{\pi}}_{d}^{(j)}-\hat{\boldsymbol{\phi}}_{d}^{(j)} & \textrm{for }d\in\bar{\mathcal{D}}
\end{array}\right.\textrm{ and }\boldsymbol{\varepsilon}_{d}^{\mathrm{dua}}\triangleq\left\{ \begin{array}{ll}
m(\tilde{\boldsymbol{\phi}}^{(j)}-\tilde{\boldsymbol{\phi}}^{(j-1)}) & \textrm{for }d=1\\
m(\hat{\boldsymbol{\phi}}_{d}^{(j)}-\hat{\boldsymbol{\phi}}_{d}^{(j-1)}) & \textrm{for }d\in\bar{\mathcal{D}}
\end{array}\right.
\]
respectively. Also, let us define the local relative tolerances as
\[
\bar{\epsilon}_{d}^{\mathrm{pri}}\triangleq\left\{ \begin{array}{ll}
\bar{\epsilon}\textrm{max}(||\tilde{\boldsymbol{\pi}}^{(j)}||_{2},||\tilde{\boldsymbol{\phi}}^{(j)}||_{2}) & \textrm{for }d=1\\
\bar{\epsilon}\mathrm{max}(||\hat{\boldsymbol{\pi}}_{d}^{(j)}||_{2},||\hat{\boldsymbol{\phi}}_{d}^{(j)}||_{2}) & \textrm{for }d\in\bar{\mathcal{D}}
\end{array}\right.\textrm{ and }\bar{\epsilon}_{d}^{\mathrm{dua}}\triangleq\left\{ \begin{array}{ll}
\bar{\epsilon}||\boldsymbol{\xi}^{(j)}||_{2} & \textrm{for }d=1\\
\bar{\epsilon}||\boldsymbol{\rho}_{d}^{(j)}||_{2} & \textrm{for }d\in\bar{\mathcal{D}}
\end{array}\right.
\]
where $\bar{\epsilon}>0$. Then the ADMM procedure terminates when
$||\boldsymbol{\varepsilon}_{d}^{\mathrm{pri}}||_{2}\leq\bar{\epsilon}_{d}^{\mathrm{pri}}$
and$||\boldsymbol{\varepsilon}_{d}^{\mathrm{dua}}||_{2}\leq\bar{\epsilon}_{d}^{\mathrm{dua}}$
for all $d\in\mathcal{D}$. This means, each MEC server check its
own stopping conditions using its local information, then notifies
the others when the stopping criteria are met. The procedure stops
when all MEC servers notify the termination.
\begin{rem}
\emph{{[}Varying penalty parameter{]}} In some cases, varying penalty
parameter might help to improve the convergence of the ADMM procedure
compared to fixed penalty parameter \cite{ADMM_boyd}. A common approach
of tuning penalty parameter is using the residual balancing scheme
\cite{ADMM_boyd}. For decentralization, we can apply the distributed
version of the scheme proposed in \cite{adaptiveADMM2016}. In particular,
let us denote by $m_{d}^{(j)}$ the penalty parameter locally used
at MEC server $d$ at iteration $j$, which are updated as 
\begin{equation}
m_{d}^{(j+1)}=\left\{ \begin{array}{ll}
m_{d}^{(j)}\tau & \textrm{for }||\boldsymbol{\varepsilon}_{d}^{\mathrm{pri}}||_{2}/\bar{\epsilon}_{d}^{\mathrm{pri}}>\beta||\boldsymbol{\varepsilon}_{d}^{\mathrm{dua}}||_{2}/\bar{\epsilon}_{d}^{\mathrm{dua}}\\
\text{\ensuremath{m_{d}^{(j)}}/\ensuremath{\tau}} & \textrm{for }||\boldsymbol{\varepsilon}_{d}^{\mathrm{dua}}||_{2}/\bar{\epsilon}_{d}^{\mathrm{dua}}>\beta||\boldsymbol{\varepsilon}_{d}^{\mathrm{pri}}||_{2}/\bar{\epsilon}_{d}^{\mathrm{pri}}\\
m_{d}^{(j)} & \textrm{otherwise}
\end{array}\right.
\end{equation}
where $\tau>1$ and $\beta>1$ are parameters. To guarantee convergence,
all $\{m_{d}^{(j)}\}_{d}$ are fixed to a predefined value after a
number of ADMM iterations. The approach might help to reach the stopping
criteria faster (numerical examples are provided in Fig.\ \ref{fig:admm_residual}).
\end{rem}
\begin{rem}
\emph{\label{remark:Maximum-admm}{[}Maximum number of ADMM iterations{]}}
One of the heuristic approaches reducing the total number of ADMM
iteration is to set the maximum number of ADMM iterations at each
InAp iteration rather than waiting until convergence to stop \cite{giangtwcom2017}.
We numerically observe that this method also has potential of accelerating
Algorithm \ref{alg:Dist_impl-small}. Numerical examples for this
method are shown in Fig.\ \ref{fig:convergeexample}.
\end{rem}

\subsubsection{Computational Complexity at the MEC Servers}

The arithmetical cost at MEC server 1 mainly comes from finding the
solution to CQP \eqref{eq:macupdate-small} in each ADMM iteration.
Subproblem \eqref{eq:macupdate-small} contains $2N(1+D+\sum_{k\in\mathcal{K}_{1}}M_{k})$
real variables, $2N$ constraints in size 3, $2N$ constraints in
size 1, $N$ constraints in size $|\mathcal{K}_{1}|(N-1)+2$ , and
one constraint in size $2M_{k}+1$ for each $k\in\mathcal{K}_{1}$.
Hence, the worst-case computational cost of using a general interior
point method for solving \eqref{eq:EE_prob-1-2-1} is $\mathcal{O}(\sqrt{5N+|\mathcal{K}_{1}|}4N^{2}(1+D+\sum_{k\in\mathcal{K}_{1}}M_{k})^{2}(10N+|\mathcal{K}_{1}|(N^{2}-N+1)+2\sum_{k\in\mathcal{K}_{1}}M_{k})$
\cite{SOCApp_boyd1999}. 

Similarly, in each ADMM iteration, MEC server $d$, $d\in\bar{\mathcal{D}}$,
solves a QCQP \eqref{eq:bsupdate-small}. The subproblem contains
$2N(1+\sum_{k\in\mathcal{K}_{d}}M_{k})$ real variables, $N$ constraints
in size $|\mathcal{K}_{d}|(N-1)+2$, $N$ constraints in size 1, and
one constraint in size $2M_{k}+1$ for each $k\in\mathcal{K}_{d}$.
So, the worst-case computational cost is $\mathcal{O}(\sqrt{2N+|\mathcal{K}_{d}|}4N^{2}(1+\sum_{k\in\mathcal{K}_{d}}M_{k})^{2}(3N+|\mathcal{K}_{d}|(N^{2}-N+1)+2\sum_{k\in\mathcal{K}_{d}}M_{k})$
\cite{SOCApp_boyd1999}. We can see that the problem solved at each
MEC server has the smaller size compared to the problem solved in
the centralized scheme, i.e.\ \eqref{eq:EE_prob-1-2-1}.

\section{Numerical Results \label{sec:Numerical-Results}}

We now numerically investigate the performance of the noncoherent
JT in dense small cell networks. We consider a circular region with
a radius of 500m centered at $\mathrm{B}_{1}$, and the small cell
BSs randomly placed in the annulus with radii $200$m and $500$m
following a uniform distribution. For the channels, both large scale
fading (path loss) and small scale fading are taken into account,
i.e., the channel vectors are modeled as $\mathbf{h}_{ik}=\sqrt{\ell_{ik}^{-\beta}}\hat{\mathbf{h}}_{ik}$,
where $\hat{\mathbf{h}}_{ik}\sim\mathcal{CN}(0,\mathbf{I})$, $\ell_{ik}$
is the distance in meters, and $\beta$ is the path loss exponent
which is taken as 5. The noise power density is $N_{0}=-174$ dBm/Hz.
We take the operation bandwidth as 1 MHz. The maximum transmission
power at the BSs are $P_{1}=40$ dBm and $P_{k}=30$ dBm, $\forall k\in\bar{\mathcal{K}}$.
The number of antennas at the BSs are $M_{1}=8$ and $M_{k}=2$, $\forall k\in\bar{\mathcal{K}}$.
The number of BSs, users, and other parameters are specified in each
experiment. 

For initial points, we randomly generate beamforming vectors $\{\mathbf{v}_{ik}\}$
so that \eqref{eq:powerconst} is satisfied. All the convex programs
in this section are solved by using the solver MOSEK \cite{Mosek}
with the modeling toolbox YALMIP \cite{YALMIP}. 

\subsection{Convergence Performance of Algorithms \ref{alg:glabalsol} and \ref{alg:IA_alg}}

\begin{figure}
\begin{centering}
\includegraphics[width=0.5\columnwidth]{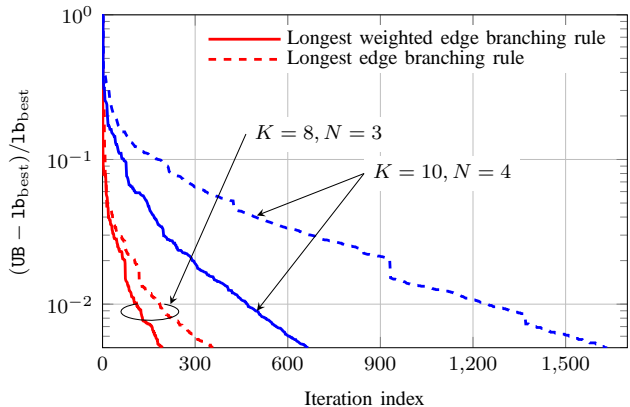}
\par\end{centering}
\caption{Convergence examples of Algorithm \ref{alg:glabalsol} with the longest
edge branching rule and longest weighted edge branching rule over
two random channel realizations corresponding to the network settings
$(K,N)=(8,3)$ and $(K,N)=(10,4)$. The weights are taken as $\mathbf{w}=[0.59;0.31;0.1]$
for $N=3$, and $\mathbf{w}=[0.097;0.519;0.135;0.249]$ for $N=4$.}
 \label{fig:convergeBnB}
\end{figure}

Fig.\  \ref{fig:convergeBnB} shows the numerical examples of convergence
of Algorithm \ref{alg:glabalsol} with the longest edge branching
rule and longest weighted edge branching rule (discussed in Remark
\ref{rem:Branchingrule}). For fairness comparison, the same initialization
of $\mathbf{r}_{\mathrm{best}}$ is used for the two branching rules
which is determined by letting $r_{i}=\log(1+\gamma_{i})$ where $\gamma_{i}$
is formed from a random feasible point $\{\mathbf{v}_{ik}\}$. In
the figure, we plot the accuracy measurement $(\mathtt{UB}-\mathtt{lb}_{\mathrm{best}})/\mathtt{lb}_{\mathrm{best}}$
(which is used as stopping criterion) as the function of the number
of iterations. We set the error tolerance parameter as $\epsilon=0.005$.
It is observed that the curves in the figure monotonically go to zero
as the number of iterations increases in all cases of channels and
branching rules. This is due to the fact that $\mathtt{lb}_{\mathrm{best}}$
monotonically increases, and the gap between $\mathtt{UB}$ and $\mathtt{lb}_{\mathrm{best}}$
monotonically decreases. Importantly, the results confirm that the
longest weighted edge branching rule can accelerate Algorithm \ref{alg:glabalsol}
compared to the conventional longest edge branching rule.

\begin{figure}
\centering
\subfigure[Convergence behavior of the considered schemes]{\label{fig:randomconvergeIA}\includegraphics[width=0.47\columnwidth]{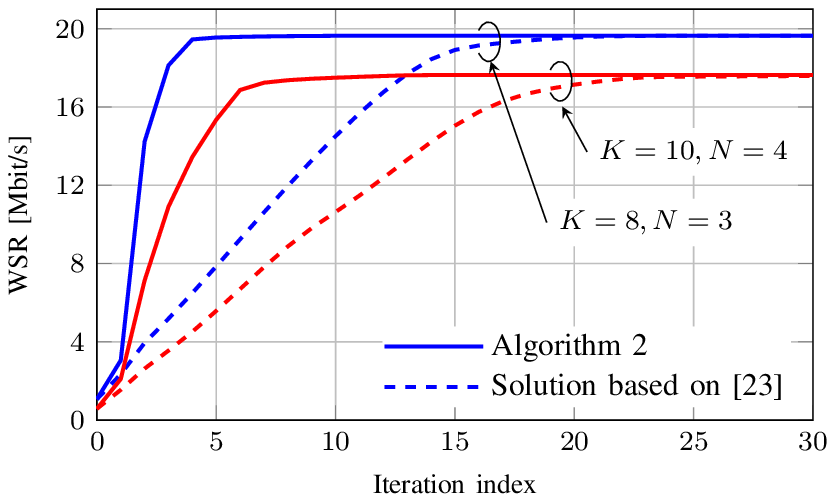}}\hspace{1mm}\subfigure[CDF of the number of iteration required to converge.]{ \label{fig:cdfconvergeIA}\includegraphics[width=0.47\columnwidth]{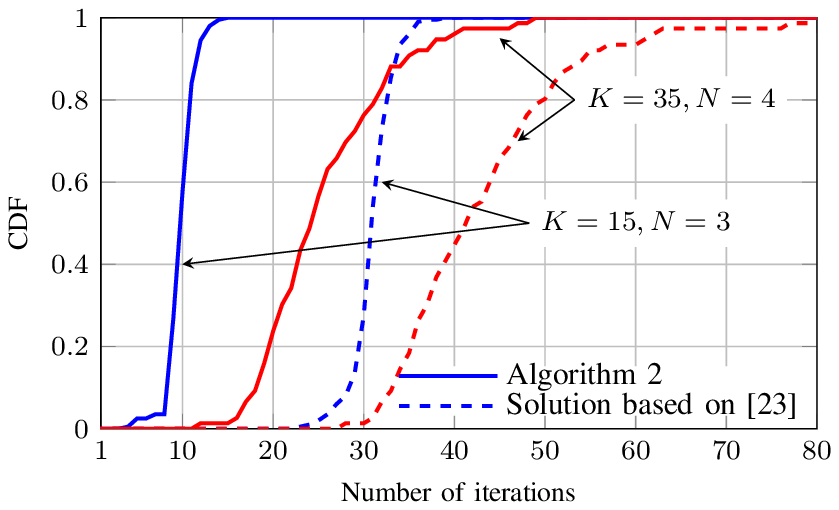}}

\caption{Convergence speed comparison of Algorithm \ref{alg:IA_alg} and the
solution modified from the one in \cite{noncoherentJSAC} with different
network settings. The weights are taken as $\mathbf{w}=[0.59;0.31;0.1]$
for $N=3$, and $\mathbf{w}=[0.097;0.519;0.135;0.249]$ for $N=4$. }
 \label{fig:convergIA}
\end{figure}
In Fig.\ \ref{fig:convergIA}, we evaluate the convergence speed
of Algorithm \ref{alg:IA_alg} compared to the solution modified from
the one in \cite{noncoherentJSAC} with different settings of $(K,N)$.
In particular, Fig.\ \ref{fig:randomconvergeIA} plots the convergence
behavior of the two schemes over two random channels realizations.
Fig.\ \ref{fig:cdfconvergeIA} plots the cumulative distribution
function (CDF) of the required number of iterations to converge. The
two schemes stop when the increase in the objective value achieved
in the last 3 iterations is less than $10^{-2}$. For each channel
realization, the same random initial point is used by both schemes
for the fairness. We can observe from the figure that the convergence
speed of Algorithm \ref{alg:IA_alg} is superior in all cases of considered
network settings. It is worth mentioning that the two schemes usually,
but not always, converge to a same value, and achieve almost the same
average performance. 

\begin{figure}[h]
\begin{centering}
\includegraphics[width=0.5\columnwidth]{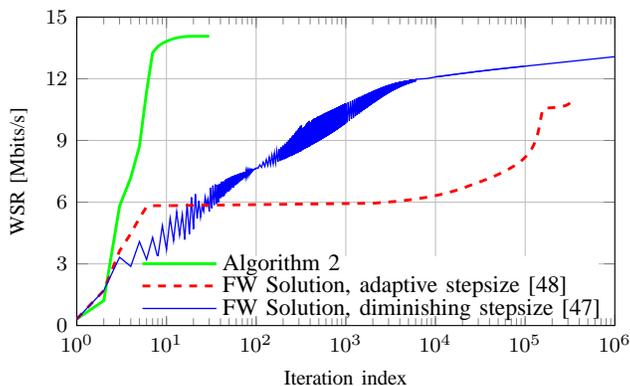}
\par\end{centering}
\caption{Convergence examples of Algorithm \ref{alg:IA_alg} in comparison
with FW solution (discussed in Remark \ref{rem:A-first-order-solution})
with two different step size rules provided in \cite{Ho_DFW_2018}
and \cite{adaptstepsizeFW} over a random channel realization corresponding
to the network setting $(K,N)=(10,4)$, $\mathbf{w}=[0.097;0.519;0.135;0.249]$.}
 \label{fig:convergeFW}
\end{figure}

In Fig.\ \ref{fig:convergeFW}, we provide an numerical example showing
the convergence behavior of the FW solution (see discussion about
FW solution in Remark \ref{rem:A-first-order-solution} and the detail
of the solution in Appendix \ref{sec:A-First-Order-Algorithm}) in
comparison with the proposed InAp-based solution. For the FW solution,
the performance of the diminishing step size rule \cite{Ho_DFW_2018}
and the adaptive step size rule \cite{adaptstepsizeFW} are provided.
The FW schemes stop when the FW gap is smaller than 1, or the number
of iteration exceeds $10^{6}$. All schemes use the same random starting
point. We can observe from the figure that, with the two step size
rules, the FW solution requires sufficiently large amount of iterations
to reach a good performance.

\subsection{Performance Comparison between Optimal Solution and Suboptimal InAp-based
Solution.}

\begin{figure}
\centering
\subfigure[Average WSR]{\label{fig:gloheur_pe}\includegraphics[width=0.47\columnwidth]{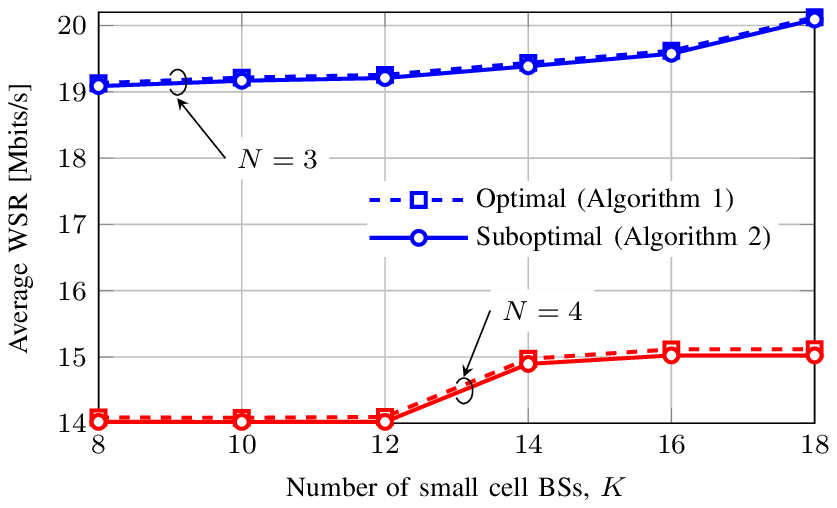}} \subfigure[CDF of the ratio of suboptimal solution to optimal solution.]{ \label{fig:gloheur_cdf}\includegraphics[width=0.47\columnwidth]{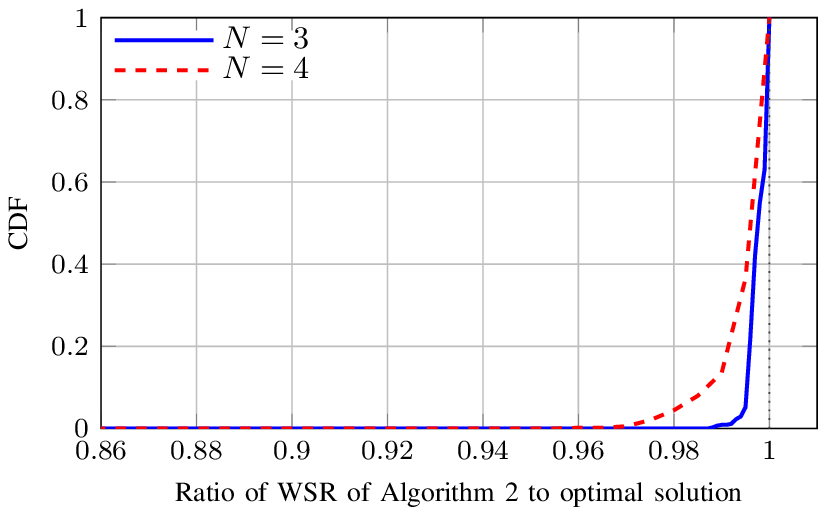}}

\caption{Performance of Algorithm \ref{alg:IA_alg} in comparison with optimal
solution (Algorithm \ref{alg:glabalsol}) with different numbers of
small cell BSs. The weights are taken as $\mathbf{w}=[0.59;0.31;0.1]$
for $N=3$, and $\mathbf{w}=[0.097;0.519;0.135;0.249]$ for $N=4$.}
 \label{fig:compaegloheur}
\end{figure}

We now evaluate the performance of Algorithm \ref{alg:IA_alg} in
terms of WSR using the globally optimal solution (Algorithm \ref{fig:convergeBnB})
as the baseline. Specifically, Fig.\ \ref{fig:gloheur_pe} plots
the average WSR performance as the function of the number of small
cell BSs. Fig.\ \ref{fig:gloheur_cdf} provides the CDF of the ratio
of WSR of Algorithm \ref{alg:IA_alg} to the optimal solution. We
can observe from Fig.\ \ref{fig:gloheur_pe} that the average WSR
performance of Algorithm \ref{alg:IA_alg} is very close to the optimal
one. In Fig.\ \ref{fig:gloheur_cdf}, we see that Algorithm \ref{alg:IA_alg}
is not worse than 96\% of the optimal solution for all channels. The
results demonstrate the efficiency of the proposed efficient solution
in terms of achieving the design objective. 

\subsection{Numerical Results of the Distributed Algorithm}

\begin{figure}
\centering{}\begin{minipage}{0.48\columnwidth}\centering\includegraphics[width=0.8\columnwidth]{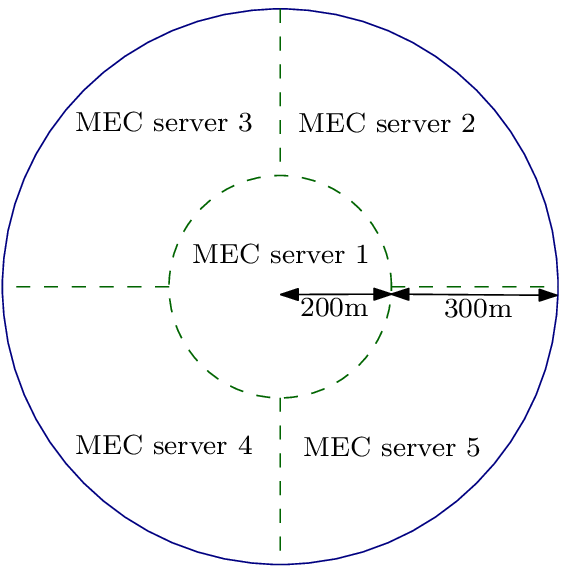}\caption{The specified serving areas of the MEC servers considered in the simulations
of decentralized networks.}
 \label{fig:silmu}\end{minipage}\hfill\begin{minipage}{0.48\columnwidth}\centering\includegraphics[width=1\columnwidth]{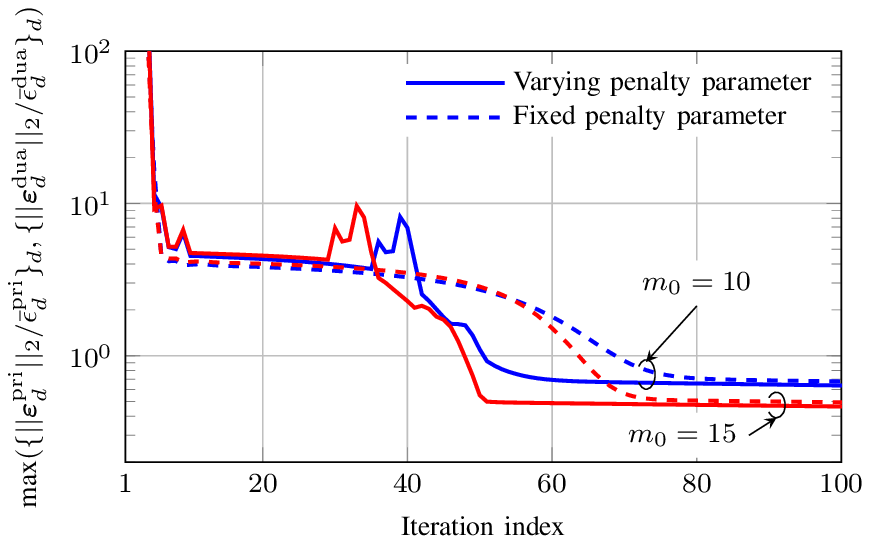}\caption{Relative residuals of fixed and varying penalty parameter schemes
over ADMM iterations at the first InAp iteration of a random channel
realization. The network setting is $(K,N)=(16,5)$. We take $w_{i}=1/N$
for all $i$.}
\label{fig:admm_residual}\end{minipage}\hfill
\end{figure}

In the next set of the experiments, we examine the performance of
Algorithm \ref{alg:Dist_impl-small}. We consider the decentralized
architecture network including 5 MEC servers. Each of them serves
the BSs lying in a specific area as shown in Fig.\ \ref{fig:silmu}.
Unless otherwise stated, we set the initial values as $q_{id}^{(0)}=y_{id}^{(0)}=1$,
for all $i,d$, $\boldsymbol{\xi}^{(0)}=\mathbf{1}$, and $\boldsymbol{\rho}_{d}^{(0)}=\mathbf{1}$
for all $d$.

Fig.\ \ref{fig:admm_residual} shows the function $\textrm{max}(\{||\boldsymbol{\varepsilon}_{d}^{\mathrm{pri}}||_{2}/\bar{\epsilon}_{d}^{\mathrm{pri}}\}_{d},\{||\boldsymbol{\varepsilon}_{d}^{\mathrm{dua}}||_{2}/\bar{\epsilon}_{d}^{\mathrm{dua}}\}_{d})$
over the ADMM iterations in the first InAp iteration of a random channel
realization. The function is introduced based on the stopping criterion,
which is satisfied when the value of the function is smaller than
1. For the adaptive penalty scheme, we take the relative tolerance
$\bar{\epsilon}$ as $10^{-3}$\cite{ADMM_boyd}. Other parameters
are taken as $\tau=2$, $\beta=5$. The initial values of penalty
parameters are set as $m_{d}=m_{0}$ for all $d$; and after 50 iterations,
they are fixed as $m_{d}=m_{0}$ for all $d$ \cite{adaptiveADMM2016}.
For the fixed penalty parameter scheme, the penalty parameter is set
as $m=m_{0}$. Value of $m_{0}$ is specified in the figure. We can
observe from the figure that, with the same chosen $m_{0}$, adaptive
(penalty) scheme can reach the stopping criteria faster.

\begin{figure}
\centering{}\centering
\subfigure[Achieved WSR]{\label{fig:admmwsr}\includegraphics[width=0.47\columnwidth]{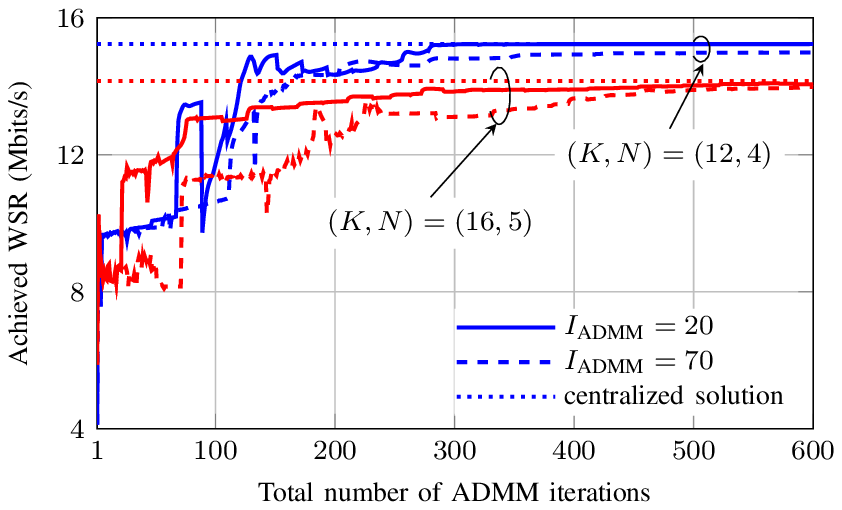}} \subfigure[Relative gap, $|f-f^{\ast}|/f^{\ast}$]{ \label{fig:admmrelativegap}\includegraphics[width=0.47\columnwidth]{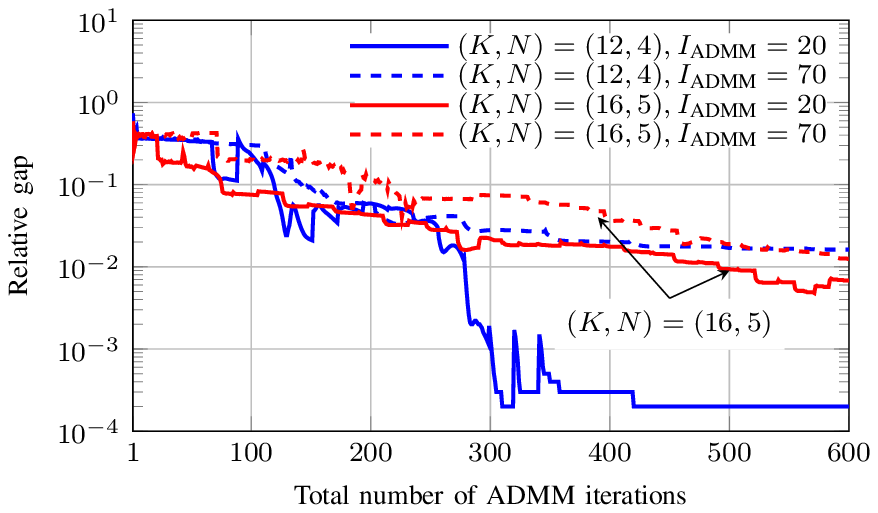}}\caption{Achieved WSR of Algorithm \ref{alg:Dist_impl-small} compared to centralized
solution over two random channel realizations corresponding to network
settings $(K,N)=(12,4)$ and $(K,N)=(16,5)$. We take $\mathbf{w}=[0.097;0.519;0.135;0.249]$
for $N=4$, and $w_{i}=1/N$, for all $i$, for $N=5$.}
 \label{fig:convergeexample}
\end{figure}

In Fig.\ \ref{fig:convergeexample}, we show the achieved WSR of
Algorithm \ref{alg:Dist_impl-small} compared to the centralized solution
(obtained by Algorithm \ref{alg:IA_alg} using the solver) over two
random channel realizations. Specifically, Fig.\ \ref{fig:admmwsr}
plots the achieved WSR as a function of the total number of ADMM iterations,
while Fig.\ \ref{fig:admmrelativegap} shows the relative gap between
the centralized and distributed solutions. To this end, let us denote
by $f$ the WSR at an ADMM iteration which is calculated by using
the beamforming vectors obtained at the iteration, and denote by $f^{\ast}$
the centralized solution. Then the relative gap is defined as $|f-f^{\ast}|/f^{\ast}$.
We take relative tolerance $\bar{\epsilon}$ as $10^{-3}$. In each
InAp iteration, the ADMM procedure stops when the termination criterion
is met or when the number of ADMM iterations exceeds $I_{\textrm{ADMM}}$.
We take $I_{\textrm{ADMM}}$ as 20 and 70. This consideration is to
illustrate the heuristic usage of the ADMM discussed in Remark \ref{remark:Maximum-admm}.
We can observe that the total number of ADMM iterations reduces remarkably
with appropriate value of $I_{\textrm{ADMM}}$. We note that the WSR
over the ADMM iterations is not necessary to be monotonic because
the ADMM works on the augmented Lagrangian function. We also note
that when the ADMM procedure converges, the sequence of WSR values
corresponding to the outer stage is monotonically increasing as Algorithm
\ref{alg:IA_alg}. However, in this figure, since $I_{\textrm{ADMM}}$
is applied, the sequence can be not monotonic since the parameters
of the outer stage are updated even when the ADMM has not converged.

\subsection{Performance Comparison between Different CoMP Strategies}

In the final set of experiments, we study the WSR performance of the
noncoherent JT in comparison with the other CoMP strategies, i.e.,
coherent JT, and CB. In CB scheme, each of users is only served by
the nearest BS. We provide the WSR performance of the noncoherent
JT based on Algorithm \ref{alg:IA_alg} due to its low complexity
and near-optimal performance. We also provide the noncoherent JT using
MRT scheme as a benchmark. The solutions of the coherent JT and CB
are derived based on that in \cite{Nam:WSRMISO:2012}.

\begin{figure}
\centering
\subfigure[$N=3$]{\label{fig:compareuser15}\includegraphics[width=0.47\columnwidth]{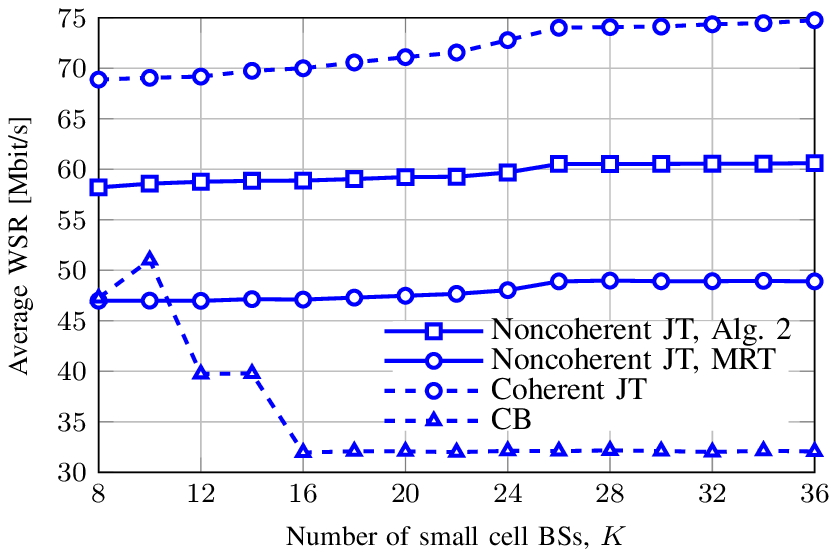}} \subfigure[$N=5$]{ \label{fig:compareuser25}\includegraphics[width=0.47\columnwidth]{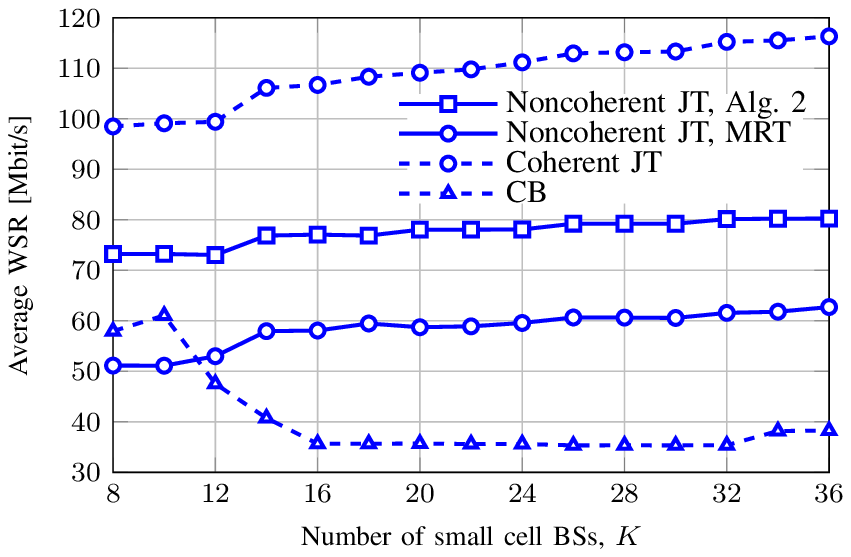}}

\caption{Average WSR performance of the considered noncoherent JT schemes,
coherent JT, and CB with different numbers of small cell BSs. The
weights are taken as $\mathbf{w}=\boldsymbol{1}$ for all cases of
$N$.}
 \label{fig:compareCOM}
\end{figure}

Fig.\ \ref{fig:compareCOM} plots the average WSR performance of
the considered schemes as functions of the number of small cell BSs.
An interesting result observed from the figure is that CB scheme might
fail to exploit the densification gain. Another result is that the
coherent JT is naturally superior to the others. Therefore, in the
networks where the coherent JT is feasible, this scheme should be
deployed to achieve maximum spectral efficiency. However, when the
synchronization accuracy is not sufficient, the noncoherent JT with
the proposed solution is a promising candidate for dense small cell
networks, since it outperforms the MRT scheme and CB, and is capable
of exploiting densification gain (the performance increases when $K$
increases).

\begin{figure}
\centering
\subfigure[$K=15$]{\label{fig:compareuser15}\includegraphics[width=0.47\columnwidth]{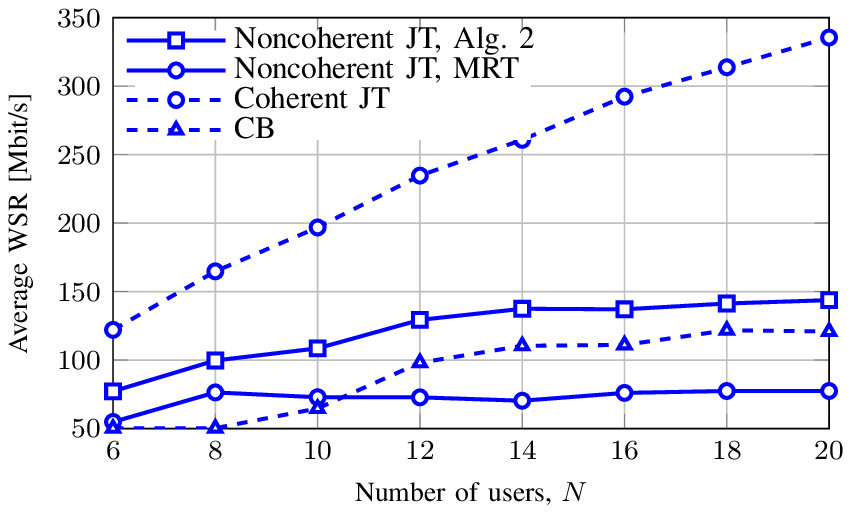}} \subfigure[$K=25$]{ \label{fig:compareuser25}\includegraphics[width=0.47\columnwidth]{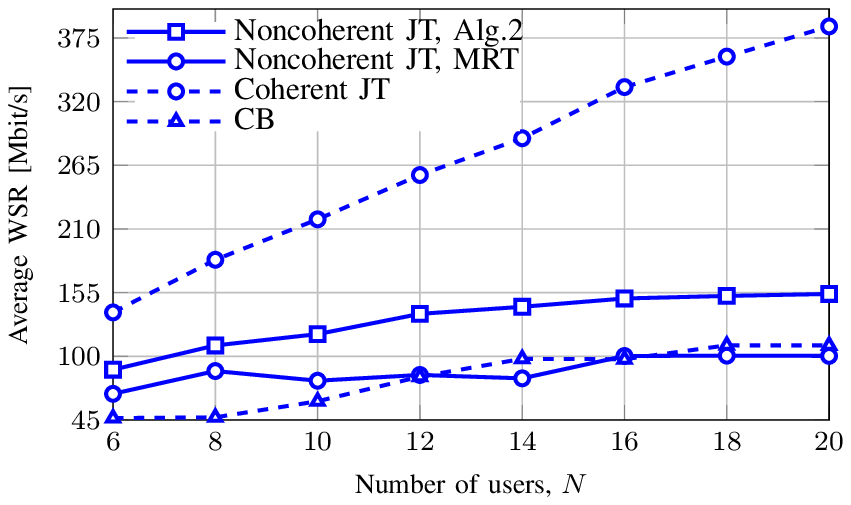}}

\caption{Average WSR performance of the considered noncoherent JT schemes,
coherent JT, and CB with different numbers of users. We take $K$
as 15 and 25. The weights are taken as $\mathbf{w}=\boldsymbol{1}$
for all cases of $N$.}
 \label{fig:compareCOM-user}
\end{figure}
In Fig.\ \ref{fig:compareCOM-user}, we show the average WSR performance
of the considered schemes as the functions of the number of users
$N$ for two cases of $K$. The main result observed from the figure
is that the WSR performance of the noncoherent JT with proposed solution,
coherent JT and CB increase when $N$ increases. This implies that
the three schemes are capable of exploiting the user diversity gain.
The results also confirm the observations taken from Fig.\ \ref{fig:compareCOM},
i.e., the coherent JT outperforms the others, and the noncoherent
JT with the proposed solution outperforms the MRT scheme and CB in
all cases of $N$. We can see that, when $N$ is large, the CB outperforms
the noncoherent JT with the MRT scheme. This might be because the
MRT scheme only considers power allocation, and thus it cannot provide
good interference management.

\section{Conclusion\label{sec:Conclusion}}

We have investigated downlink noncoherent JT in dense small cell networks.
Particularly, we have considered the problem of designing beamforming
vectors at the macro cell and small cell BSs for maximizing WSR. Because
the problem is intractable, we have developed a BRnB algorithm to
achieve globally optimal solution. In addition, for practical implementation,
we have developed a low-complexity algorithm based on the IA optimization
framework, which has been numerically shown to be able to achieve
near optimal performance. Moreover, in order to implement the InAp-based
solution on decentralized networks using MEC servers, we have provided
a distributed algorithm based on the ADMM. The results have revealed
that noncoherent JT is capable of exploiting densitification gain
and outperforming the CB. It is also feasible to implement the transmission
scheme distributively, and it does not require as strict synchronization
accuracy as the coherent JT. Thus, noncoherent JT is a promising transmission
scheme for dense small cell networks in terms of the WSR performance.

\appendices{}

\section{Proof of Lemmas}

\subsection{Proof of Lemma \ref{lem:feasibility}\label{sec:Proof-of-Lemma-feasibility}}

The proof is based on that of \cite[Theorem 1]{Small_cell_MassMIMO}.
The if part is obvious since $\tilde{\mathcal{S}}_{\mathrm{SD}}(\mathbf{r})$
is the SDR of $\tilde{\mathcal{S}}(\mathbf{r})$ achieved by removing
rank-one constraints. Thus we focus on the only-if part, i.e., if
$\tilde{\mathcal{S}}_{\mathrm{SD}}(\mathbf{r})$ is nonempty, $\tilde{\mathcal{S}}(\mathbf{r})$
is nonempty. If there exists $\mathbf{V}'{}_{ik}\in\tilde{\mathcal{S}}_{\mathrm{SD}}(\mathbf{r})$
such that $\mathrm{rank}(\mathbf{V}_{ik})\leq1,$$\forall i\in\mathcal{N},\forall k\in\mathcal{K}$,
then $\tilde{\mathcal{S}}(\mathbf{r})$ is nonempty which completes
the proof. Now suppose that there exists $\mathbf{V}'_{uv}\in\tilde{\mathcal{S}}_{\mathrm{SD}}(\mathbf{r})$
where $\mathrm{rank}(\mathbf{V}'_{uv})>1$ for some $(u,v)$. Then
consider the problem \begin{subequations}\label{eq:feasible_prob_sd-1}
\begin{align}
\underset{\mathbf{V}}{\maxi} & \;\mathbf{h}_{uv}\mathbf{V}\mathbf{h}_{uv}\herm\label{eq:GeneralObj-2-1-1-1-1}\\
\st & \;\mathbf{h}_{jv}\mathbf{V}\mathbf{h}_{jv}\herm\leq\mathbf{h}_{jv}\mathbf{V}'_{uv}\mathbf{h}_{jv}\herm,\forall j\neq u\label{eq:relaxation1-1}\\
 & \;\tr(\mathbf{V})\leq\tr(\mathbf{V}'_{uv}),\mathbf{V}\in\mathbb{S}_{+}^{M_{v}}.\label{eq:relaxation2-1}
\end{align}
\end{subequations}It is proved based on a primal-dual analysis that
\eqref{eq:feasible_prob_sd-1} always has a rank-one solution (c.f
\cite[Appendix III]{ZF:wie:2008}), which completes the proof.

\subsection{Proof of Lemma \ref{thm:global} \label{sec:Proof-of-Theorem}}

The proof is based on the convergence analysis of BRnB in \cite{tuy2005monotonic,robustBjon}.
Particularly, the reduction is valid, i.e., no feasible point in a
box having better performance than $\mathtt{lb}_{\mathrm{best}}$
is lost after the reduction. The bounding guarantees that the upper
bound of a box is non-increasing and $\mathtt{lb}_{\mathrm{best}}$
is non-decreasing. Thus the gap $\mathtt{UB}-\mathtt{lb}_{\mathrm{best}}$
is monotonically decreasing. In addition, the bisection partition
in the branching is exhaustive \cite{tuy2005monotonic}. These lead
to $(\mathtt{UB}-\mathtt{lb}_{\mathrm{best}})\rightarrow0$ as $t\rightarrow\infty$,
i.e., the gap of the bounds uniformly converges to zero. This completes
the proof.

\subsection{Proof of Lemma \ref{lem:equivalentIA} \label{subsec:Proof-of-eqiIA}}

To prove the first statement, we show that: (i) the constraints in
\eqref{eq:snrconstr} hold with equality at optimality, and (ii) vectors
$\{\mathbf{v}_{ik}^{\star}\}$ achieve optimal value of (\ref{eq:wsr_prob}).
For (i), suppose that \eqref{eq:snrconstr} is inactive at the optimality
for some $i$, then there exists $\mu'_{i}>\mu_{i}^{\star}$ such
that constraint $i$ in \eqref{eq:snrconstr} holds with equality.
Thus, the point with $\mu'_{i}$ instead of $\mu_{i}^{\star}$ is
still feasible and results in a strictly larger objective value, since
$\log(1+\mu'_{i})>\log(1+\mu_{i}^{\star})$. This contradicts with
the assumption that an optimal solution has been achieved. 

For (ii), suppose that there exists a feasible point of (\ref{eq:wsr_prob})
denoted by $\{\mathbf{v}'_{ik}\}$ such that ${\textstyle \sum_{i\in\mathcal{N}}}w_{i}\log(1+\gamma_{i}(\{\mathbf{v}'_{ik}\}))>{\textstyle \sum_{i\in\mathcal{N}}}w_{i}\log(1+\mu_{i}^{\star})$,
then we determine $\mu'_{i}=\gamma_{i}(\{\mathbf{v}'_{ik}\})$. Clearly,
$(\{\mathbf{v}'_{ik}\},\{\mu'_{i}\})$ is a feasible point of (\ref{eq:EE_prob-1})
which achieves larger objective value than $(\{\mathbf{v}_{ik}^{\star}\},\{\mu_{i}^{\star}\})$.
Again, this contradicts with the assumption that $(\{\mathbf{v}_{ik}^{\star}\},\{\mu_{i}^{\star}\})$
is optimal. 

Similarly, for the converse, we can show by contradiction that there
exist no feasible points of (\ref{eq:EE_prob-1}) which can achieve
better objective value than $(\{\mathbf{v}_{ik}^{\ast}\},\{\gamma_{i}(\{\mathbf{v}_{ik}^{\ast}\})\})$.
Finally, the the final statement follows the fact (i).

\subsection{Proof of Lemma \ref{lem:ConvergeIA} \label{subsec:Proof-ConvergeIA}}

By contradiction, we can justify that at optimality all constraints
in \eqref{eq:log_appr} hold with equality, leading to $\sqrt{1+\mu_{i}^{(t)}}=1/\pi_{i}^{(t)}$
$\forall\,i,t$. Let us consider iteration $t$. According to the
principles of the InAp, the solution to \eqref{eq:EE_prob-1-2-1}
in iteration $t$ is feasible to the problem in iteration $(t+1)$,
and thus ${\textstyle \sum_{i\in\mathcal{N}}}\tilde{w}_{i}^{(t)}\pi_{i}^{(t)}\geq{\textstyle \sum_{i\in\mathcal{N}}}\tilde{w}_{i}^{(t)}\pi_{i}^{(t+1)}$
which is equivalent to
\begin{align*}
{\textstyle \sum_{i\in\mathcal{N}}}w_{i}\sqrt{(1+\mu_{i}^{(t)})/(1+\mu_{i}^{(t)})} & \geq{\textstyle \sum_{i\in\mathcal{N}}}w_{i}\sqrt{(1+\mu_{i}^{(t)})/(1+\mu_{i}^{(t+1)})}\\
\Leftrightarrow{\textstyle \sum_{i\in\mathcal{N}}}\underset{=\log(1+\mu_{i}^{(t)})}{w_{i}\underbrace{(J^{(t)}-2\sqrt{(1+\mu_{i}^{(t)})/(1+\mu_{i}^{(t)})})}} & \leq{\textstyle \sum_{i\in\mathcal{N}}}w_{i}\underset{\leq\log(1+\mu_{i}^{(t+1)})}{\underbrace{(J^{(t)}-2\sqrt{(1+\mu_{i}^{(t)})/(1+\mu_{i}^{(t+1)})})}}
\end{align*}
where $J^{(t)}=\log(1+\mu_{i}^{(t)})+2$. In addition, problem (\ref{eq:EE_prob-1})
is upper bounded by a finite value due to power constraints \eqref{eq:powerconst-1}.
This completes the proof.

\subsection{Proof of Lemma \ref{lem:convergeADMM}\label{sec:Proof-of-ConvergeADMM}}

The proof is based on the convergence analysis of ADMM in \cite{ADMM_boyd}.
Firstly, we note that the feasible set of \eqref{eq:EE_prob-1-2-1}
is convex and nonempty for all InAp iterations, and so is that of
\eqref{eq:EE_prob-1-1-1}. The objective of \eqref{eq:EE_prob-1-2-1}
is bounded. Consequently, sets $\tilde{\mathcal{S}}$ and $\hat{\mathcal{S}}_{d}$
are nonempty, and the problems \eqref{eq:macupdate-small}, \eqref{eq:bsupdate-small},
and \eqref{eq:globalupdate-small} are solvable \cite[Assumption 1]{ADMM_boyd}.
Secondly, we recall that the considered problem is only constrained
by the maximum transmit power at the BSs. Thus, problem \eqref{eq:EE_prob-1-1-1}
is strictly feasible. Consequently, the assumption that the unaugmented
Lagrangian (i.e., function \eqref{eq:lagrangian-small} with $m=0$)
has a saddle point holds \cite[Assumption 2]{ADMM_boyd}. With these,
the lemma follows the statement in \cite[Section 3.2.1]{ADMM_boyd}.

\section{A First-Order Algorithm Solving \eqref{eq:wsr_prob} via Conditional
Gradient Method\label{sec:A-First-Order-Algorithm}}

In this appendix, we present a first-order solution for solving \eqref{eq:wsr_prob}
using Frank-Wolfe's method. To proceed, let us define some notations
as $\mathbf{H}_{ik}=\mathbf{h}_{ik}\herm\mathbf{h}_{ik}$, $\hat{\mathbf{H}}_{ik}=\textrm{blkdiag}\{\underset{N}{\underbrace{\mathbf{H}_{ik},...,\mathbf{H}_{ik}}}\}$,
$\mathbf{H}_{i}=\textrm{blkdiag}\{\hat{\mathbf{H}}_{i1},...,\hat{\mathbf{H}}_{i(k+1)}\}$,
$\mathbf{v}_{k}=[\mathbf{v}_{1k};...;\mathbf{v}_{Nk}]$, $\mathbf{v}=[\mathbf{v}_{1};...;\mathbf{v}_{k+1}]$,
$\hat{\mathbf{G}}_{ik}=\textrm{blkdiag}\{\underset{i-1}{\underbrace{\mathbf{H}_{ik},...,\mathbf{H}_{ik}}},\linebreak\boldsymbol{0},\underset{N-i}{\underbrace{\mathbf{H}_{ik},...,\mathbf{H}_{ik}}}\}$,
and $\mathbf{G}_{i}=\textrm{blkdiag}\{\hat{\mathbf{G}}_{i1},...,\hat{\mathbf{G}}_{i(k+1)}\}$.
Then we rewrite the objective function as 
\[
g(\mathbf{v})={\textstyle \sum_{i\in\mathcal{N}}}w_{i}\log\Bigl(\frac{\mathbf{v}\herm\mathbf{H}_{i}\mathbf{v}+\sigma_{i}^{2}}{\mathbf{v}\herm\mathbf{G}_{i}\mathbf{v}+\sigma_{i}^{2}}\Bigr)
\]
For ease of exposition, we convert $g(\mathbf{v})$ in the real-domain
as 
\[
g(\tilde{\mathbf{v}})={\textstyle \sum_{i\in\mathcal{N}}}w_{i}\log\Bigl(\frac{\tilde{\mathbf{v}}\trans\bar{\mathbf{H}}_{i}\tilde{\mathbf{v}}+\sigma_{i}^{2}}{\tilde{\mathbf{v}}\trans\bar{\mathbf{G}}_{i}\tilde{\mathbf{v}}+\sigma_{i}^{2}}\Bigr)
\]
where $\tilde{\mathbf{v}}=[\Re\{\mathbf{v}\};\Im\{\mathbf{v}\}]$,
$\bar{\mathbf{H}}_{i}=[\Re\{\mathbf{H}_{i}\},-\Im\{\mathbf{H}_{i}\};\Im\{\mathbf{H}_{i}\},\Re\{\mathbf{H}_{i}\}]$,
and $\bar{\mathbf{G}}_{i}=[\Re\{\mathbf{G}_{i}\},\linebreak-\Im\{\mathbf{G}_{i}\};\Im\{\mathbf{G}_{i}\},\Re\{\mathbf{G}_{i}\}]$.
Let $\tilde{\mathbf{v}}^{(t)}$ be a feasible point, then the gradient
of $g(\tilde{\mathbf{v}})$ at $\tilde{\mathbf{v}}^{(t)}$ is 
\begin{eqnarray}
\nabla_{\tilde{\mathbf{v}}}g(\tilde{\mathbf{v}}^{(t)}) & = & {\textstyle \sum_{i\in\mathcal{N}}}w_{i}\Bigl(\frac{2\bar{\mathbf{H}}_{i}\tilde{\mathbf{v}}^{(t)}}{(\tilde{\mathbf{v}}^{(t)})\trans\bar{\mathbf{H}}_{i}\tilde{\mathbf{v}}^{(t)}+\sigma_{i}^{2}}-\frac{2\bar{\mathbf{G}}_{i}\tilde{\mathbf{v}}^{(t)}}{(\tilde{\mathbf{v}}^{(t)})\trans\bar{\mathbf{G}}_{i}\tilde{\mathbf{v}}^{(t)}+\sigma_{i}^{2}}\Bigr).\label{eq:gradientg}
\end{eqnarray}
Then the linear optimization oracle at each iteration is 
\begin{equation}
\tilde{\mathbf{v}}^{\ast}=\textrm{argmax}(\nabla g(\tilde{\mathbf{v}}^{(t)})\trans\tilde{\mathbf{v}}|\tilde{\mathbf{v}}_{k}\trans\tilde{\mathbf{v}}_{k}\leq P_{k},\forall k\in\mathcal{K})\label{eq:linear_O}
\end{equation}
 where $\tilde{\mathbf{v}}_{k}=[\Re\{\mathbf{v}_{k}\};\Im\{\mathbf{v}_{k}\}]$.
As the objective and the feasible set of the above problem are separable
with respect to $\tilde{\mathbf{v}}_{k}$, it is easy to see that
\eqref{eq:linear_O} has the following closed-form expression 
\begin{equation}
\tilde{\mathbf{v}}_{k}^{\ast}=\sqrt{P_{k}/(\mathbf{c}_{k}\trans\mathbf{c}_{k})}\mathbf{c}_{k}\label{eq:FW-closeform}
\end{equation}
 where $\mathbf{c}_{k}=\{\nabla g(\tilde{\mathbf{v}}^{(t)})\}_{\tilde{\mathbf{v}}_{k}}$
is the vector including the elements of $\nabla g(\tilde{\mathbf{v}}^{(t)})$
associated with $\tilde{\mathbf{v}}_{k}$.

The first-order iterative algorithm for solving \eqref{eq:wsr_prob}
is outlined as follows:

\textbf{Initialization}: Set small $\epsilon_{\mathrm{G}}$, $t\coloneqq1$,
choose initial point $\tilde{\mathbf{v}}^{(1)}$.

\textbf{repeat}

\hspace{4mm}Generate $\nabla g(\tilde{\mathbf{v}}^{(t)})$, then
determine $\tilde{\mathbf{v}}^{\ast}$ according to \eqref{eq:FW-closeform}.

\hspace{4mm}\textbf{if} $\nabla g(\tilde{\mathbf{v}}^{(t)})\trans(\tilde{\mathbf{v}}^{\ast}-\tilde{\mathbf{v}}^{(t)})\leq\epsilon_{\mathrm{G}}$

\hspace{4mm}\hspace{4mm}Stop and return $\tilde{\mathbf{v}}^{(t)}$

\hspace{4mm}\textbf{else}

\hspace{4mm}\hspace{4mm}Choose step size $\alpha^{(t)}\in[0,1]$,
then update $\tilde{\mathbf{v}}^{(t+1)}:=(1-\alpha^{(t)})\tilde{\mathbf{v}}^{(t)}+\alpha^{(t)}\tilde{\mathbf{v}}^{\ast}$
\label{FW:move}

\hspace{4mm}\hspace{4mm}$t:=t+1$.

\noindent \textbf{\indent }\hspace{4mm}\textbf{end}

\noindent In each iteration, the new iterate $\tilde{\mathbf{v}}^{(t+1)}$
is determined by moving the current iterate $\tilde{\mathbf{v}}^{(t)}$
along the dictated direction $\tilde{\mathbf{v}}^{\ast}-\tilde{\mathbf{v}}^{(t)}$
with step size $\alpha^{(t)}\in(0,1]$.

\subsection*{Smoothness of $g(\mathbf{v})$}

We now investigate the smoothness of $g(\tilde{\mathbf{v}})$, which
is essential for the convergence result of the Frank-Wolfe (FW) based
method above to be provided in the next subsection. In particular,
the Hessian of $g(\mathbf{v})$ at $\tilde{\mathbf{v}}^{(t)}$ is
given by
\begin{multline}
\nabla_{\tilde{\mathbf{v}}}^{2}g(\tilde{\mathbf{v}}^{(t)})={\textstyle \sum_{i\in\mathcal{N}}}w_{i}\Bigl(\frac{2\bar{\mathbf{H}}_{i}}{(\tilde{\mathbf{v}}^{(t)})\trans\bar{\mathbf{H}}_{i}\tilde{\mathbf{v}}^{(t)}+\sigma_{i}^{2}}-\frac{4\bar{\mathbf{H}}_{i}\tilde{\mathbf{v}}^{(t)}(\tilde{\mathbf{v}}^{(t)})\trans\bar{\mathbf{H}}_{i}\trans}{((\tilde{\mathbf{v}}^{(t)})\trans\bar{\mathbf{H}}_{i}\tilde{\mathbf{v}}^{(t)}+\sigma_{i}^{2})^{2}}\\
-\frac{2\bar{\mathbf{G}}_{i}}{(\tilde{\mathbf{v}}^{(t)})\trans\bar{\mathbf{G}}_{i}\tilde{\mathbf{v}}^{(t)}+\sigma_{i}^{2}}+\frac{4\bar{\mathbf{G}}_{i}\tilde{\mathbf{v}}^{(t)}(\tilde{\mathbf{v}}^{(t)})\trans\bar{\mathbf{G}}_{i}\trans}{((\tilde{\mathbf{v}}^{(t)})\trans\bar{\mathbf{G}}_{i}\tilde{\mathbf{v}}^{(t)}+\sigma_{i}^{2})^{2}}\Bigr).\label{eq:hessianG-1}
\end{multline}
It is trivial to show that $||\nabla_{\tilde{\mathbf{v}}}^{2}g(\tilde{\mathbf{v}}^{(t)})||_{2}\leq{\displaystyle \rho_{\mathrm{L}}\triangleq}\sum_{i\in\mathcal{N}}w_{i}(\frac{2}{\sigma_{i}^{2}}||\bar{\mathbf{H}}_{i}||_{2}+\frac{4P_{\mathrm{total}}}{\sigma_{i}^{4}}||\bar{\mathbf{H}}_{i}||_{2}^{2}+\frac{2}{\sigma_{i}^{2}}||\bar{\mathbf{G}}_{i}||_{2}+\frac{4P_{\mathrm{total}}}{\sigma_{i}^{4}}||\bar{\mathbf{G}}_{i}||_{2}^{2})$,
where $P_{\mathrm{total}}=\sum_{k}P_{k}$, which means $g(\tilde{\mathbf{v}})$
is a $\rho_{\mathrm{L}}$-smooth function \cite{lecture_gradient}.

\subsection*{Step Size Rules and Convergence}

Note that the FW-type method for nonconvex problems is not monotonically
increasing, and thus a proper choice of the step size $\alpha^{(t)}$
at each iteration is critical for the convergence. This issue is relatively
open and receiving increasing interest but there are some known step
size rules for the conditional gradient method. One is the line-search
rule: $\alpha^{(t)}=\textrm{argmax}(g((1-\alpha)\tilde{\mathbf{v}}^{(t)}+\tilde{\mathbf{v}}^{\ast})|\alpha^{(t)}\in[0,1])$
\cite{convergenonconvexFW}, but it is not cost effective here since
the one-dimentional search does not admit a closed-form solution.
An adaptive step size rule is recently proposed in \cite{adaptstepsizeFW}.
The diminishing step size rule $\alpha^{(t)}=t^{-\omega}$, where
$\omega\in(0.5,1]$, can also be used \cite{Ho_DFW_2018}.

For the convergence, we recall that the feasible set of problem \eqref{eq:wsr_prob}
is convex and compact. In addition, we have shown that $g(\tilde{\mathbf{v}})$
has a finite Lipschitz gradient constant $\rho_{\mathrm{L}}$. Thus,
by the mentioned step size rules, it is guaranteed that the iterative
procedure converges to a stationary point of \eqref{eq:wsr_prob}
\cite{convergenonconvexFW,Ho_DFW_2018,adaptstepsizeFW}. In addition,
for the line-search and adaptive step size rules, it is proved that
the convergence rate is $\mathcal{O}(1/\sqrt{t})$ \cite[Theorem 1]{convergenonconvexFW},
\cite[Theorem 1]{adaptstepsizeFW}.

\bibliographystyle{IEEEtran}
\bibliography{IEEEabrv,References}

\end{document}